\documentclass[12pt]{iopart}
\usepackage{epsfig}
\usepackage{amsfonts}
\usepackage{color}

\begin{document}

\newtheorem{df}{Definition}
\newtheorem{remark}{Remark}

\title[Multi-winged Lorenz attractors]{Multi-winged Lorenz attractors due to bifurcations of a periodic orbit with multipliers $(-1,i,-i)$}

\author{Efrosiniia Karatetskaia$^1$, Alexey Kazakov$^{1,*}$, Klim Safonov$^{1}$ \& Dmitry Turaev$^{2}$ }

\address{$^1$ National Research University Higher School of Economics, \\ 25/12 Bolshaya Pecherskaya Ulitsa, 603155 Nizhny Novgorod, Russia}
\address{$^2$ Imperial College, London SW7 2AZ, United Kingdom}
\ead{eyukaratetskaya@gmail.com, kazakovdz@yandex.ru, safonov.klim@yandex.ru and d.turaev@imperial.ac.uk}

\vspace{10pt}
\begin{indented}
\item[]April 2024
\end{indented}

\begin{abstract}
We show that bifurcations of periodic orbits with multipliers $(-1,i,-i)$ can lead to the birth of pseudohyperbolic (i.e., robustly chaotic) Lorenz-like attractors of three different types: one is a discrete analogue of the classical Lorenz attractor, and the other two are new. We call them two- and four-winged ``Sim\'o angels''. These three attractors exist in an orientation-reversing, three-dimensional, quadratic H\'enon map. Our analysis is based on a numerical study of a normal form for this bifurcation, a three-dimensional system of differential equations with a Z4-symmetry. We investigate bifurcations in the normal form and describe those responsible for the emergence of the Lorenz attractor and the continuous-time version of the Simo angels. Both for the normal form and the 3D H\'enon map, we have found open regions in the parameter space where the attractors are pseudohyperbolic, implying that for every parameter value from these regions every orbit in the attractor has positive top Lyapunov exponent.
\end{abstract}

%
\vspace{2pc}
\noindent{\it Keywords}: normal forms, Lorenz attractor, pseudohyperbolicity.
%
%
%
%

\section{Introduction}

A remarkable discovery by Arneodo, Coullet, Spiegel, and Tresser \cite{ACS83, ACT85, ACST85} is that the triple instability of an equilibrium state or a periodic orbit can lead to chaos. The reason is that local bifurcations corresponding to three zero Lyapunov exponents are described by normal forms which are nonlinear systems of differential equations in $\mathbb{R}^3$. Numerous scenarios for the emergence of chaotic dynamics in such systems are known, see e.g. \cite{Sh80, ACT82, Sh86, IR05}. The most celebrated examples of normal forms with chaotic behavior are the Arneodo-Coullet-Spiegel-Tresser models \cite{ACT85, ACST85}, Morioka-Shimizu system \cite{ASh86, SM80, ASh93}, Lorenz model \cite{Lor63, SST93}. They have interesting and rich dynamics and, because the phase space dimension is not high, they can be studied in depth. This is particularly important, because the knowledge of the properties of chaos in a normal form can be transferred to any system undergoing the local bifurcation that is described by this normal form. In this way, systems of absolutely different physical nature can be shown to exhibit the same universal patterns of chaotic behavior.

All strange attractors can be divided into two groups: genuinely chaotic attractors and quasiattractors. Each orbit of an attractor of the first type has positive maximal Lyapunov exponent and, most importantly, this property persists under small perturbations. Therefore, such attractors demonstrate chaotic dynamics robustly. The other type of strange attractors, the quasiattractors, either contain stable periodic orbits (with extremely narrow, so elusive in experiments, absorbing domains) or such orbits appear inside the attractor under arbitrarily small perturbations~\cite{AfrSh83a}. Thus, one can never be sure wether a chaotic attractor is observed or it is just a transient dynamics, and finally orbits would converge to a simple attractor. A natural question arises here: how to distinguish genuinely chaotic attractors from qusiattractors?

An answer to this question was given in \cite{GKT21} where ``P or Q'' conjecture was proposed. According to this conjecture a property of an attractor to be genuine is equivalent to its pseudohyperbolicity -- a weak version of hyperbolicity. A theory of pseudohyperbolic attractors was developed in \cite{TS98, TS08}. The common property of pseudohyperbolic attractors is the existence of a continuous splitting of the tangent space in a neighborhood of an attractor into a direct sum of two invariant linear subspaces $E^{ss}$ and $E^{cu}$: the linearized system restricted to $E^{ss}$ is uniformly contracting, whereas in $E^{cu}$ it uniformly expands volumes. In the case of classical hyperbolic attractors, the field of subspaces $E^{cu}$ is given by tangents to the unstable manifolds. In the case of pseudohyperbolicity, the linearized system can contract some directions in $E^{cu}$, however, the total volume expansion in this subspace ensures that the maximal Lyapunov exponent for every orbit of the attractor is always positive. It is important that any possible contraction in $E^{cu}$ should be uniformly weaker than any contraction in $E^{ss}$. This property guarantees that both the positivity of the maximal Lyapunov exponent and the whole pseudohyperbolic structure are preserved at small perturbations of the system.

In \cite{TS98, TS08} it was also shown that the class of pseudohyperbolic attractors, besides the classical hyperbolic attractors, also includes singular-hyperbolic (Lorenz) attractors and ``wild'' attractors which contain periodic orbits with homoclinic tangencies. Both these types of attractors are not structurally stable due to bifurcations occurring inside them when parameters change (the appearance of homoclinic loops in the case of singular-hyperbolic attractors and inevitable homoclinic and heteroclinic tangencies between invariant manifolds of saddle periodic orbits in the case of wild attractors). However these bifurcations do not produce stable periodic orbits, only saddles occurs inside the attractor.

\begin{figure}[tb]
\center{\includegraphics[width=1.0\linewidth]{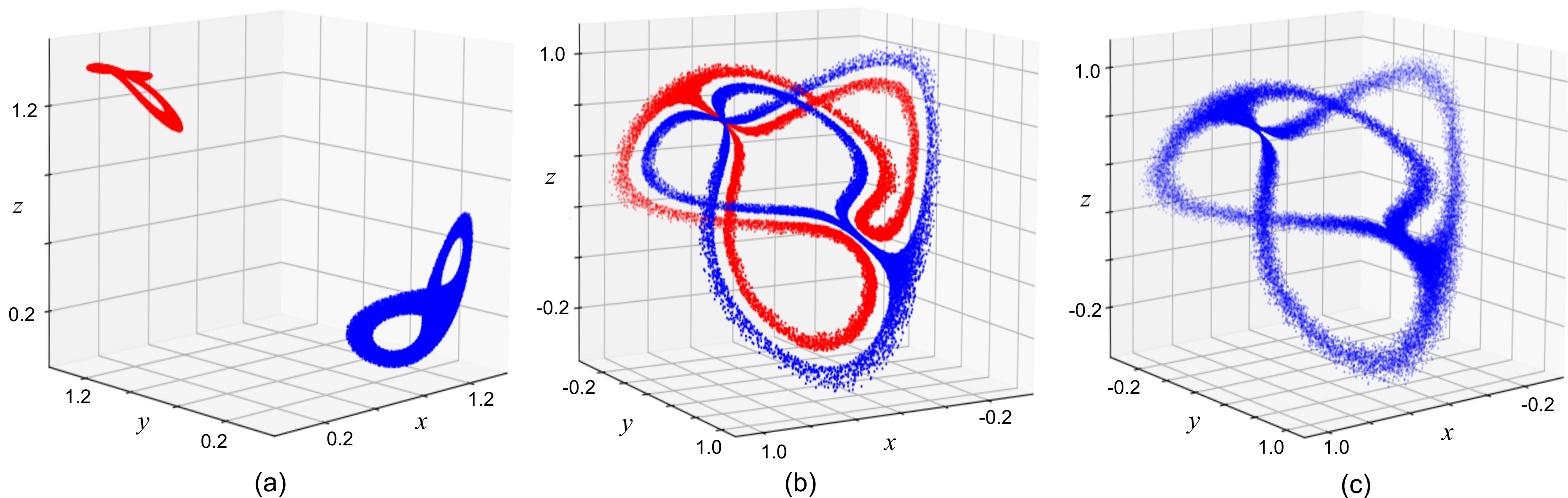} }
\caption{{\footnotesize Pseudohyperbolic attractors in map \eref{eq_HenonMap}: (a) period-2 Lorenz attractor, $(B, M_1, M_2) = (-0.854, 2.783, -1.25)$; (b) two-winged Simo angel, $(B, M_1, M_2) = (-0.877, 1.703, -0.85)$ (c) four-winged Simo angel, $(B, M_1, M_2) = (-0.875, 1.7, -0.85)$. In (a) and (b) red and blue are two components of the same attractor (the map takes red component to blue and blue -- to red).}}
\label{fig0a}
\end{figure}

In this paper, we show that bifurcations of periodic orbits with multipliers $(-1,i,-i)$ can lead to the birth of wild pseudohyperbolic Lorenz-like attractors of three different types. The first type is a discrete analogue of the classical Lorenz attractor. This attractor contains a saddle point of period 2, see Fig.~\ref{fig0a}a. The other two types of attractors are new. The attractor of second type also has two separated components, see Fig.~\ref{fig0a}b. When these two components merge we observe an attractor of third type, see Fig.~\ref{fig0a}c. We note that this attractor was first observed numerically by Carles Sim\'o in \cite{GOST05}. Therefore, we call the attractors of second and third types as two-winged and four-winged Simo angels. Formal definitions for all three types of attractors are given in Sec.~\ref{sec2}.

We find these attractors in an orientation-reversing, quadratic, three-dimensional H\'enon map
\begin{equation}
\left\{
\begin{array}{l}
\bar x = y, \\
\bar y = z, \\
\dot z = M_1 + B x + M_2 y - z^2
\end{array}
\right.
\label{eq_HenonMap}
\end{equation}
with parameters $M_1, M_2$, and $B$. The Jacobian of this map is equal to $B$. Since we consider the orientation-reversing case, $B$ is always negative. This map has up to 2 fixed points. At $(B, M_1, M_2) = (-1, 7/4, -1)$ one of its fixed point $P(x,y,z): x=y=z=(B+M_2-1 + \sqrt{(B+M_2-1)^2+4 M_1})/2$ has multipliers $(-1,i,-i)$. We find that two regions with pseudohyperbolic attractors adjoint this codimension-3 point: in one region we find a period-2 discrete Lorenz attractor, see Fig.~\ref{fig0a}a; while in another one, depending on parameters, we observe the two- or four-winged attractors presented in Fig.~\ref{fig0a}b and Fig.~\ref{fig0a}c. All these attractors inevitably contain orbits of homoclinic tangencies and, as a result, wild hyperbolic sets \cite{N70, N74, N79, GST91, GST93a, GST93b, PV94}. Therefore we say that they are wild.

In \cite{KKST24} we derive a normal form for bifurcations of a periodic orbit with multipliers $(-1,i,-i)$. In the case of map \eref{eq_HenonMap} this normal form is given by
\begin{equation}
\left\{
\begin{array}{l}
\dot x = -\alpha x - \beta y - \frac{1}{2} z (x - y) + \frac{1}{2} (x^2 y + x y^2) + \frac{3}{8} z^2 x - \frac{1}{8} z^2 y, \\
\dot y = \beta x - \alpha y + \frac{1}{2} z (x + y) + \frac{1}{2} (x^2 y - x y^2) + \frac{1}{8} z^2 x + \frac{3}{8} z^2 y, \\
\dot z = \mu z + x^2 - y^2 - \frac{1}{4} z^3 - \frac{1}{2} z (x^2 + y^2),
\end{array}
\right.
\label{eq_mainEq}
\end{equation}
where $\alpha, \beta$, and $\mu$ are small parameters. Note that this system possesses a Z4-symmetry
\begin{equation}
\mathcal{S}: x \rightarrow y, \quad y \rightarrow -x, \quad z \rightarrow -z.
\label{eq_Z4sym}
\end{equation}
This is a normal form for map \eref{eq_HenonMap} in the sense that in the neighborhood of the fixed point $P$ for $(B, M_1, M_2)$ close to the critical value $(-1, 7/4, -1)$ the fourth iteration of this map, after some change of coordinates and parameters, becomes close to the time-1 map of system \eref{eq_mainEq} composed with $\mathcal S$.

For each values of parameters, system \eref{eq_mainEq} has the trivial equilibrium $O(0,0,0)$. When $\mu > 0$, it has an additional $\mathcal S$-symmetric pair of equilibria $O^+(0,0,2\sqrt{\mu})$ and $O^-(0,0,-2\sqrt{\mu})$. The equilibrium $O$ corresponds to the fixed point $P$ of map \eref{eq_HenonMap}, and the pair $(O^+,O^-)$ corresponds to an orbit of period-2.

\begin{figure}[tb]
\center{\includegraphics[width=1.0\linewidth]{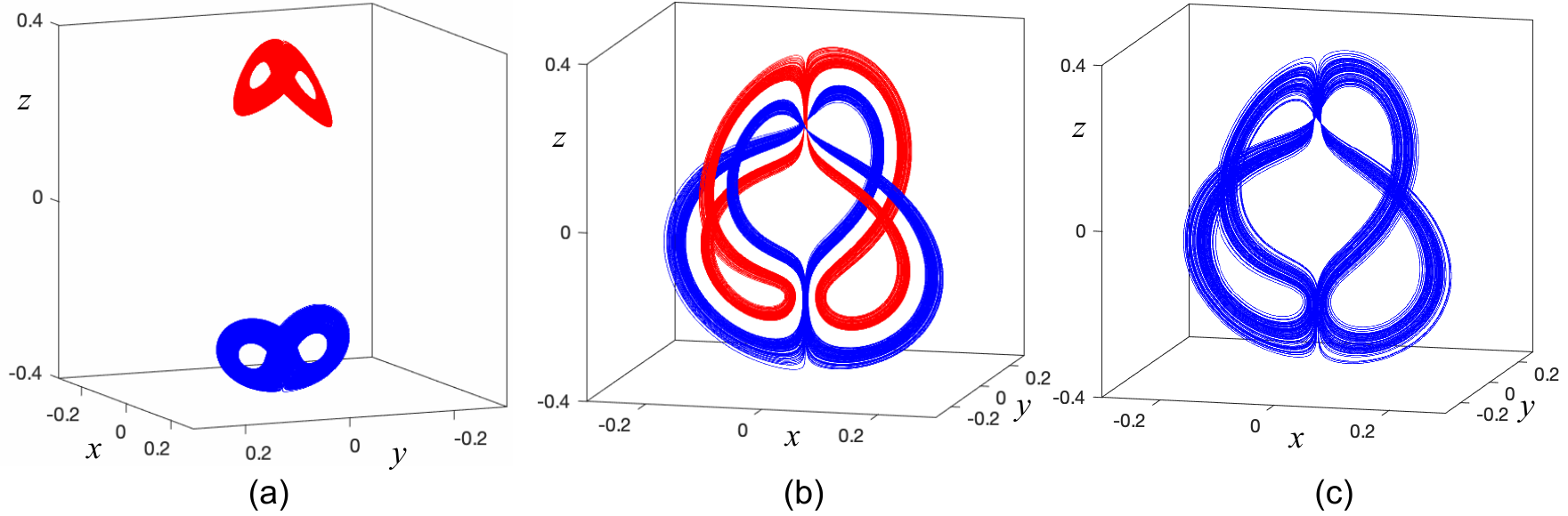} }
\caption{{\footnotesize Pseudohyperbolic attractors in system \eref{eq_mainEq}: (a) a pair of Lorenz attractors, $(\alpha, \beta, \mu) = (0.1, 0.2, 0.04)$; (b) two-winged Simo angel, $(\alpha, \beta, \mu) = (0.08, 0.004, 0.04)$; (c) four-winged Simo angel, $(\alpha, \beta, \mu) = (0.08, 0.006, 0.04)$.}}
\label{fig0b}
\end{figure}

In \cite{KKST24} it is proven that the normal form \eref{eq_mainEq} has parameter regions where there exist continuous-time analogues of the discrete period-2 Lorenz attractor and the Simo angels, see Fig.~\ref{fig0b}. In the present paper, we perform a numerical analysis of the normal form \eref{eq_mainEq}, describe main bifurcations leading to the formation of these attractors, and determine the boundaries of their existence regions in the parameter space.

We apply the results to the study of map~\eref{eq_HenonMap} and find the regions of existence of the discrete Lorenz attractors and Simo angels near the bifurcation of the point $P$ with multipliers $(-1, i, -i)$. We check numerically the pseudohyperbolicity (hence -- robust chaoticity) of these attractors.

Note that the three-dimensional H\'enon map~\eref{eq_HenonMap} is the universal approximation for the rescaled first-return map near a homoclinic tangency of the class where dynamics are not reducible to two-dimensional maps~\cite{GMO06}. Chaotic dynamics in such a map has been studied in \cite{GGTat07, GO13, GGOT13, GOT14, GGKS21, HamptonMeiss2022_1, HamptonMeiss2022_2}. Since robustly chaotic attractors persist at small perturbations, it follows that the new types of attractors we have found in the H\'enon map~\eref{eq_HenonMap} emerge at bifurcations of arbitrary homoclinic tangencies of this class.

The paper is organized as follows. In Section~\ref{sec2} we explain what is meant by the pseudohyperbolicity. We also present geometric models for attractors shown in Fig.~\ref{fig0b} and give formal definitions for them. Section~\ref{sec3} is devoted to the study of pseudohyperbolic attractors in the normal form~\eref{eq_mainEq}. We show numerically that these attractors satisfy the conditions given in Sec.~\ref{sec2}, we also determine the bifurcations leading to each type of attractors and describe their boundaries of existence. In Section~\ref{sec3} we transfer these results to the case of the 3D H\'enon map~\eref{eq_HenonMap}.

\section{Main definitions} \label{sec2}

The Lorenz attractor was discovered in \cite{Lor63}. Nowadays there exist different definitions of Lorenz-like attractors \cite{G76, ABS77, GW79, W79, ABS82, MPP04, Pujals07, PujalsSambarino2009, BBP21}. In this paper, by a Lorenz-like attractor we mean a chaotic attractor of a system of ODEs whose Poincar\'e map satisfies the conditions similar to the conditions of the Afraimovich-Bykov-Shilnikov geometric model \cite{ABS77, ABS82}. To formulate these conditions we use the notion of pseudohyperbolicity introduced in \cite{TS98, TS08}, see also \cite{OT17, CTZ18, GKT21}.

\begin{df} A system of ODEs or a diffeomorhism $F$, given in $\mathbb{R}^n$, is pseudohyperbolic in a compact forward-invariant set ${\cal A} \subseteq \mathbb{R}^n$ if it possesses the following properties.
\begin{itemize}
\item[{\rm 1)}]
At each point $x \in {\cal A}$ there exist two continuously dependent on $x$ linear subspaces, $E_1(x)$ with $\dim E_1 = k$ and $E_2(x)$ with $\dim E_2 = n-k$, which are invariant with respect to the differential $DF$ of the system, i.e., for all $t \geq 0$
$$
DF_t E_1(x) = E_1(F_t(x)), \qquad DF_t E_2(x) = E_2(F_t(x)),
$$
where $F_t$ is the time-$t$ map if $F$ is the system of ODEs, and it is the $t$-th iteration of $F$ if $F$ is a diffeomorhism.
\item[{\rm 2)}] The splitting to $E_1$ and $E_2$ is dominated, i.e., there exist constants $C_1 > 0$ and $\beta > 0$ such that
$$
\|DF_t(x)|_{E_2}\| \cdot \|(DF_t(x)|_{E_1})^{-1}\| \leq C_1 e^{-\beta t}
$$
for all $t \geq 0$ and $x \in {\cal A}$.
\item[{\rm 3)}] The differential $DF$ in $E_1$ stretches all $k$-dimensional volumes exponentially, i.e., there exist constants $C_2 > 0$ and $\sigma > 0$ such that
$$
det(DF_t(x)|_{E_1}) \geq C_2 e^{\sigma t}
$$
for all $t \geq 0$ and $x \in {\cal A}$.
\end{itemize}
\label{df_ph}
\end{df}

The first two conditions guarantee that the decomposition of the tangent space of the system to the invariant continuous subspaces $E_1$ and $E_2$ is robust with respect to small perturbations. The third condition guaranties the positiveness of the maximal Lyapunov exponent for each forward orbit in ${\cal A}$.

If the forward-invariant set ${\cal A}$ is an attractor, we say that the attractor is pseudohyperbolic. Following~\cite{TS98, GT17}, we use the Ruelle's definition of the attractor~\cite{Ruelle81}:
\begin{df}
An attractor is a stable, compact, chain-transitive invariant set.
\label{df_ph2}
\end{df}
We call a set \textit{stable} if for any $\delta > 0$ there exists $\varepsilon > 0$ such that $\varepsilon$-orbits that start at this set never run away from its $\delta$-neighborhood. The set is \textit{chain-transitive} if any its two point are connected by an $\varepsilon$-orbit for every $\varepsilon > 0$.

\subsection{Geometric model for the two- and four-winged attractors}

For simplicity, we further only consider three-dimensional systems of ODEs $\dot X = F(X)$. Assume that this system:
\begin{itemize}
\item[(A1)] possesses Z4-symmetry $\mathcal{S}$, as in \eref{eq_Z4sym};
\item[(A2)] has a pair of equilibria $O^+$ and $O^-$ such that $O^+ = \mathcal{S}(O^-)$ and $O^- = \mathcal{S}(O^+)$, and eigenvalues $\gamma, \lambda_1$, and $\lambda_2$ of these equilibria satisfy the conditions: $\lambda_2 < \lambda_1 < 0 < \gamma$ and $-\lambda_1 / \gamma < 1$, i.e., $O^+$ and $O^-$ are pseudohyperbolic with $\dim E_1 = 2$ and $\dim E_2 = 1$;
\item[(A3)] is pseudohyperbolic with $\dim E_1 = 2$ and $\dim E_2 = 1$ in an open forward-invariant domain $\mathcal D$ that contains $O^\pm$.
\end{itemize}

\begin{figure}[tb]
\center{\includegraphics[width=1.0\linewidth]{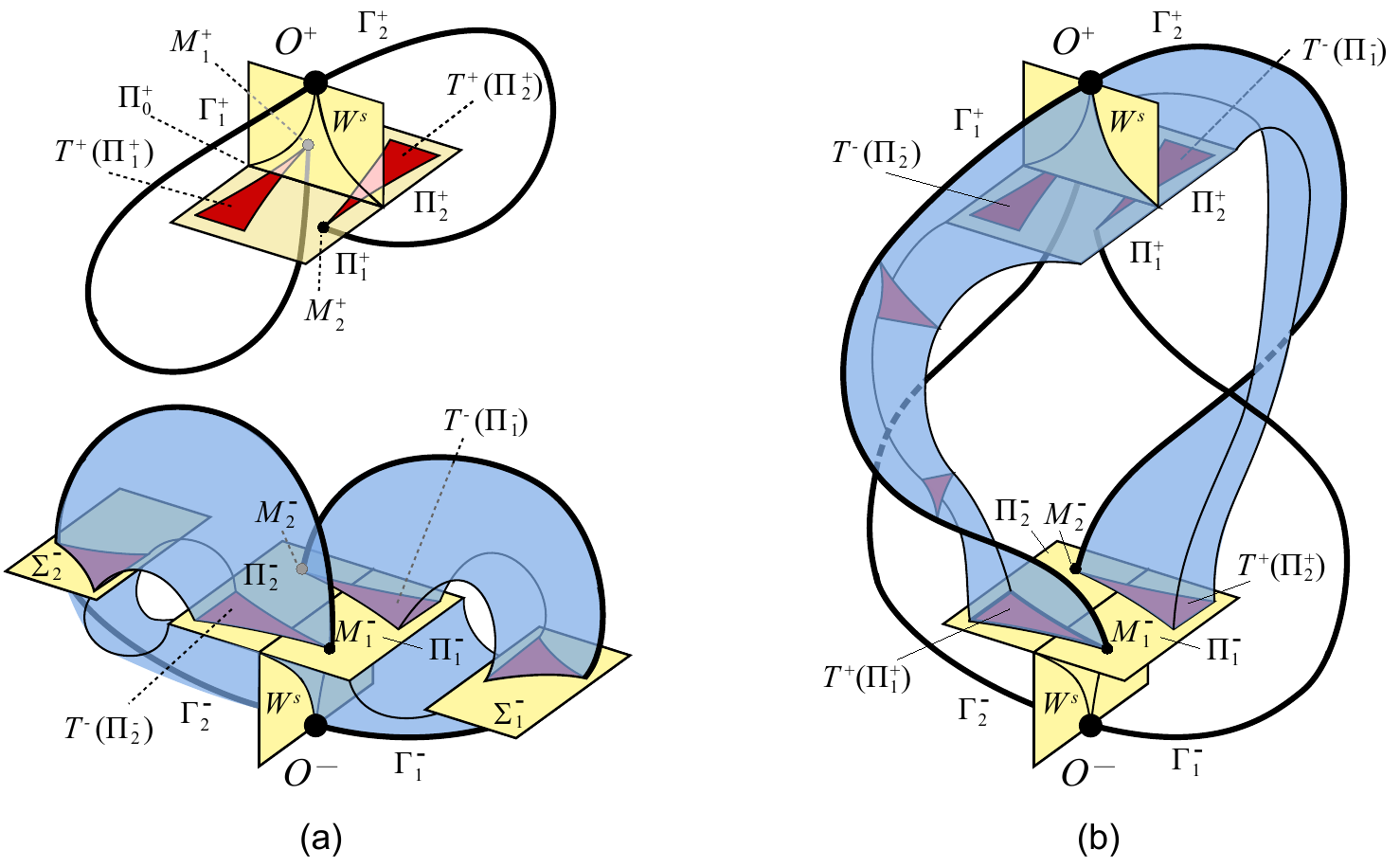} }
\caption{{\footnotesize Geometric models for (a) the Lorenz attractor and (b) the Simo angels: $O^+$ and $O^-$ are a pair of $\mathcal{S}$-symmetric equilibria ($O^+ = \mathcal{S}(O^-)$ and $O^- = \mathcal{S}(O^+)$) with two-dimensional stable ($W^s$) and one-dimensional unstable ($W^u$) invariant manifolds. $\Gamma^\pm_{1,2}$ are the unstable separatrices forming together with $O^\pm$ the unstable invariant manifold of these equilibria. $T^+$ and $T^-$ are maps from $\Pi^+$ to $\Pi^+$ and from $\Pi^-$ to $\Pi^-$ in the case (a), and from $\Pi^+$ to $\Pi^-$ and from $\Pi^-$ to $\Pi^+$ in the case (b).}}
\label{fig_GM}
\end{figure}

Then, we introduce two cross-sections $\Pi^+$ and $\Pi^-$ that lie in $\mathcal D$ and are transversal to the two-dimensional stable manifolds $W^s(O^+)$ and $W^s(O^-)$, respectively, see Figure~\ref{fig_GM}. $\Pi^+$ ($\Pi^-$) is divided by $W^s(O^+)$ ($W^s(O^-)$) into two parts $\Pi^+_1$ and $\Pi^+_2$ ($\Pi^-_1$ and $\Pi^-_2$).

\vspace{0.4cm}

Assume that:
\begin{itemize}
\item[(A4)] every forward orbit in $\mathcal D$ either belongs to $W^s(O^+) \cup W^s(O^-)$ or intersects $\Pi^+_1$ and $\Pi^+_2$ (or both);
\item[(A5)] every forward orbit starting in $\Pi^+_1 \cup \Pi^+_2$ or in $\Pi^-_1 \cup \Pi^-_2$ returns to $\Pi^+ \cup \Pi^-$.
\end{itemize}

It follows from~\cite{TS08} that the pseudohyperbolicity and the existence of the cross-sections $\Pi^+$ and $\Pi^-$ imply that the stable manifold $W^s(O^+) \cup W^s(O^-)$ is dense in $\mathcal D$. Therefore, for any point $Q \in \mathcal D$ at least one of the two equilibria $O^+$ and $O^-$ is attainable by $\varepsilon$-orbits from $Q$. This implies that there can be no more than two attractors in $\mathcal D$: the attractor $\mathcal{A}^+$ is the set of all points attainable from $O^+$ by $\varepsilon$-orbits for every $\varepsilon > 0$, and $\mathcal{A}^-$ is the set of all points attainable from $O^-$ by $\varepsilon$-orbits for every $\varepsilon > 0$. Note that if $O^+$ is attainable from $O^-$ and (by the symmetry) $O^-$ is attainable from $O^+$, then $\mathcal{A}^+ = \mathcal{A}^-$ is the only attractor in $\mathcal D$.

Conditions A4, A5 imply that the dynamics of the system in $\mathcal D$ are defined by the Poincar\'e map $\Pi^+ \cup \Pi^- \to \Pi^+ \cup \Pi^-$. Here we can have two cases: the orbits starting at $\Pi^+_1 \cup \Pi^+_2$ return to $\Pi^+$ without an intersection with $\Pi^-$ (as in Fig.~\ref{fig_GM}a), or they first hit $\Pi^-$ (as in Fig.~\ref{fig_GM}b). By the symmetry, in the first case the orbits starting at $\Pi^-_1 \cup \Pi^-_2$ return to $\Pi^-$ without an intersection with $\Pi^+$, and in the second case they hit $\Pi^+$ first. To distinguish between these two cases, one looks at the behavior of the one-dimensional unstable manifolds of $O^+$ and $O^-$. The unstable manifold $W^u(O^+)$ consists of two separatrices $\Gamma^+_1$ and $\Gamma^+_2$, $W^u(O^-)$ consists of two separatrices $\Gamma^-_1$ and $\Gamma^-_2$; we have the first case when $\Gamma^+_{1,2}$ return to $\Pi^+$ without intersecting $\Pi^-$ (by the symmetry, $\Gamma^-_{1,2}$ return to $\Pi^-$ without intersecting $\Pi^+$), and in the second case $\Gamma^+_{1,2}$ intersect with $\Pi^-$ first (by the symmetry, $\Gamma^-_{1,2}$ intersect with $\Pi^+$ first). See an illustration in Fig.~\ref{fig_GM}.

In the first case, the absorbing domain $\mathcal{D}$ consists of two connected components: one contains the equilibrium $O^+$ and the orbits which intersect $\Pi^+$, and another contains the equilibrium $O^-$ and the orbits which intersect $\Pi^-$. Thus, we have two disjoint attractors $\mathcal{A}^+$ and $\mathcal{A}^-$, and we will say that the system $\dot X=F(X)$ has a \textit{symmetric pair of Lorenz attractors} in this case.

In the second case, the domain $\mathcal D$ is connected and we say that the system $\dot{X}=F(X)$ has a \textit{Simo angel attractor}. We have two possibilities in this case.
\begin{itemize}
\item The equilibria $O^+$ and $O^-$ belong to two different chain-transitive sets. In this case we say that the system has a \textit{pair of two-winged Simo angels} $\mathcal{A}^+$ and $\mathcal{A}^-$.
\item The equilibria $O^+$ and $O^-$ belong to the same chain-transitive set $\mathcal{A}^+ = \mathcal{A}^-$. In this case we have a \textit{four-winged Simo angel} which contains both equilibria.
\end{itemize}

The pseudohyperbolicity implies that the Poincar\'e return map to $\Pi^+$ or $\Pi^-$ is hyperbolic: it has an invariant foliation $F^+_{ss}$ in $\Pi^+$ and $F^-_{ss}$ in $\Pi^-$ such that the map is contracting along the leaves of the foliation and expanding in the direction transverse to the leaves. Thus, in the case of a pair of Lorenz attractors we have in each of the two components of $\mathcal{D}$ the classical Afraimovich-Bykov-Shilniov geometric model, see Fig.~\ref{fig_GM}a. We denote as $T^+$ the map acting from $\Pi^+$ to $\Pi^+$, and as $T^-$ the map acting from $\Pi^-$ to $\Pi^-$. The map $T^+$ has a discontinuity line $\Pi^+_0 = W^s(O^+) \cap \Pi^+$. All orbits starting at $\Pi^+_0$ tend to $O^+$ as $t \to +\infty$. All other orbits in $\Pi^+$ reach the cross-section $\Pi^+$ in a finite time. By continuity, we define the image $T^+(\Pi^+_0)$ by the points $M^+_1$ and $M^+_2$ of the first intersection with $\Pi^+$ of the separatrices $\Gamma^+_1$ and $\Gamma^+_2$, respectively, see Fig.~\ref{fig_GM}a. Thus, the images $T^+(\Pi^+_1)$ and $T^+(\Pi^+_2)$ are ``wedges'' with the end points $M^+_1$ and $M^+_2$. By the symmetry, we have the same picture on the cross-section $\Pi^-$.

A similar geometric model for the Simo angel is shown in Fig.~\ref{fig_GM}b. We denote as $T^+$ the map from $\Pi^+$ to $\Pi^-$, and as $T^-$ the map from $\Pi^-$ to $\Pi^+$. As before, the lines $\Pi^+_0 = W^s(O^+) \cap \Pi^+$ and $\Pi^-_0 = W^s(O^-) \cap \Pi^-$ are the discontinuities of $T^+$ and $T^-$, respectively. The images $T^+(\Pi^+_1)$ and $T^+(\Pi^+_2)$ are ``wedges'' in $\Pi_-$ with the end points $M^-_1$ and $M^-_2$, and the images $T^-(\Pi^-_1)$ and $T^-(\Pi^-_2)$ are ``wedges'' in $\Pi_+$ with the end points $M^+_1$ and $M^+_2$, see Fig.~\ref{fig_GM}b.

In the Simo angel case, due to the symmetry \eref{eq_Z4sym}, the map $T^-$ is conjugate to $T^+$:
$$
T^- = \mathcal{S}^{-1} \circ T^+ \circ \mathcal{S}.
$$
This means that the \textit{first-return map} $T = T^- \circ T^+$ from $\Pi^+$ to $\Pi^+$ is given by
\begin{equation}
T = -(\mathcal{S} \circ T^+)^2.
\label{eq_ST2}
\end{equation}
Indeed, this follows since $T = T^- \circ T^+ = \mathcal{S}^{-2} \circ \mathcal{S} \circ T^+ \circ \mathcal{S} \circ T^+$ and $\mathcal{S} \circ \mathcal{S} = -id$.

We call $\hat T^+ = \mathcal{S} \circ T^+: \Pi^+_1 \cup \Pi^+_2 \to \Pi^-$ the \textit{half-return map}. Due to the symmetry we have
$$
\hat T^+ \circ (-id) = (-id) \circ \hat T^+.
$$
By~\eref{eq_ST2}, the dynamics of the first-return map $T$ is completely captured by the iterations of the half-return map $\hat T^+$.

\begin{figure}[tb]
\center{\includegraphics[width=0.9\linewidth]{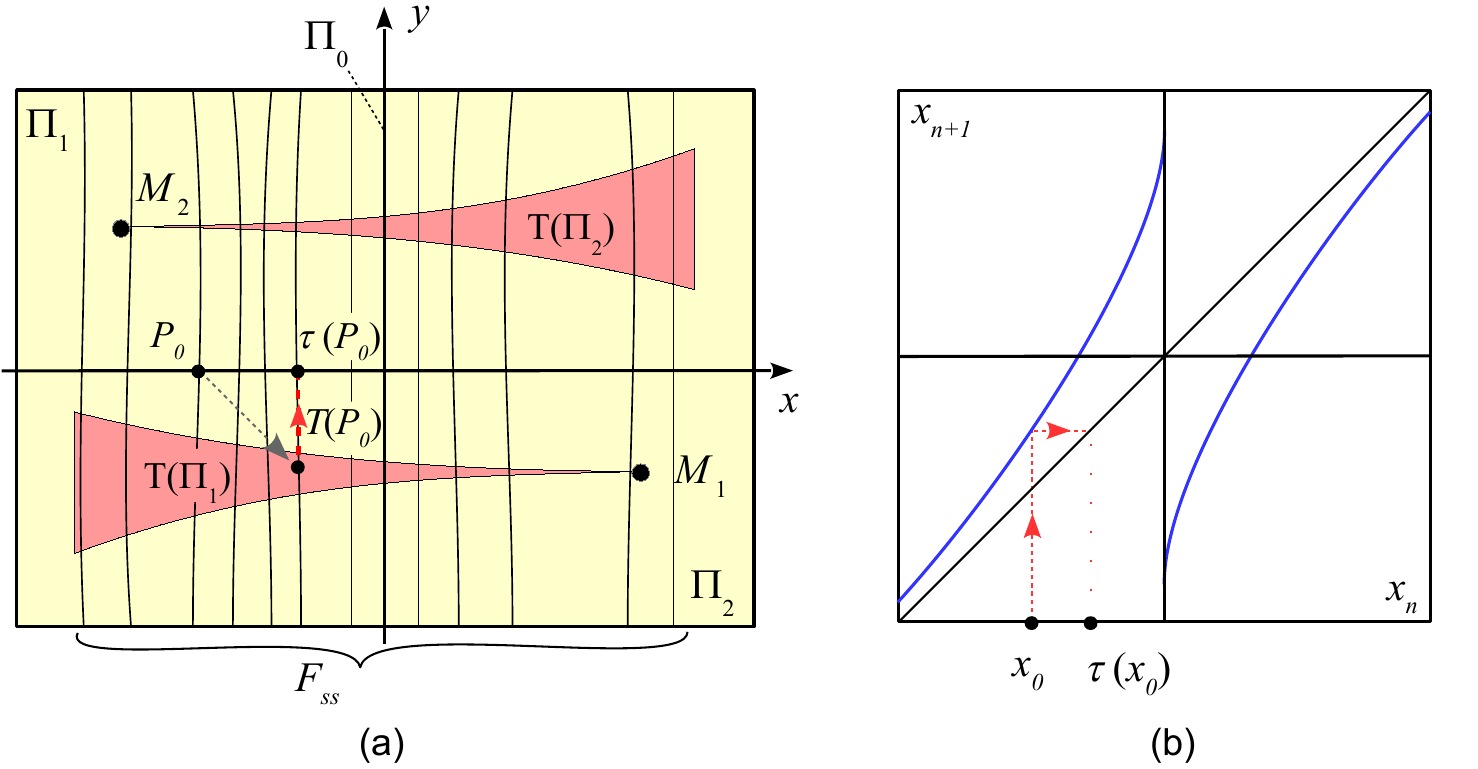} }
\caption{{\footnotesize (a) Strong stable invariant foliation $F_{ss}$ in $\Pi$; (b) quotient map $\tau$ by the leaves of $F_{ss}$.}}
\label{fig_QM}
\end{figure}

Note that the half-return map $\hat T^+$ has the same properties as the map $T^+$ in the case of the pair of Lorenz attractors and is also described by the Afraimovich-Bykov-Shilnikov geometric model. In particular, the pseudohyperbolicity implies that one can introduce coordinates on $\Pi^+$ such that the conditions (*) in \cite{ABS77} or (1.1) in \cite{ABS82} are fulfilled. As we mentioned, this implies the existence of the strong stable invariant foliation $F^+_{ss}$ which contains the line $\Pi^+_0$ and all its preimages, see Fig.~\ref{fig_QM}a. These preimages fill the cross-section $\Pi^+$ densely \cite{ABS82}. The existence of the invariant foliation $F^+_{ss}$ allows to consider the quotient map -- a piece-wise smooth, expanding, one-dimensional map whose dynamics completely determine the dynamics of $\hat T^+$ ($T^+$) \cite{Mal85}, see Fig.~\ref{fig_QM}b. In particular, the structure of both Lorenz attractors and Simo angels are completely determined by a kneading-invariant of the quotient map (an infinite symbolic sequence which codes the behavior of the unstable separatrices, see Sec.~\ref{sec_3_4}).

Note an interesting detail about the Simo angels. The minus sign in formula~\eref{eq_ST2} implies that the first-return map $T$ changes orientation in the strong-stable foliation $F^+_{ss}$. Therefore, the Simo angels should be classified as non-orientable Lorenz-like attractors in the terminology of \cite{ABS82}.

\begin{figure}[tb]
\center{\includegraphics[width=0.8\linewidth]{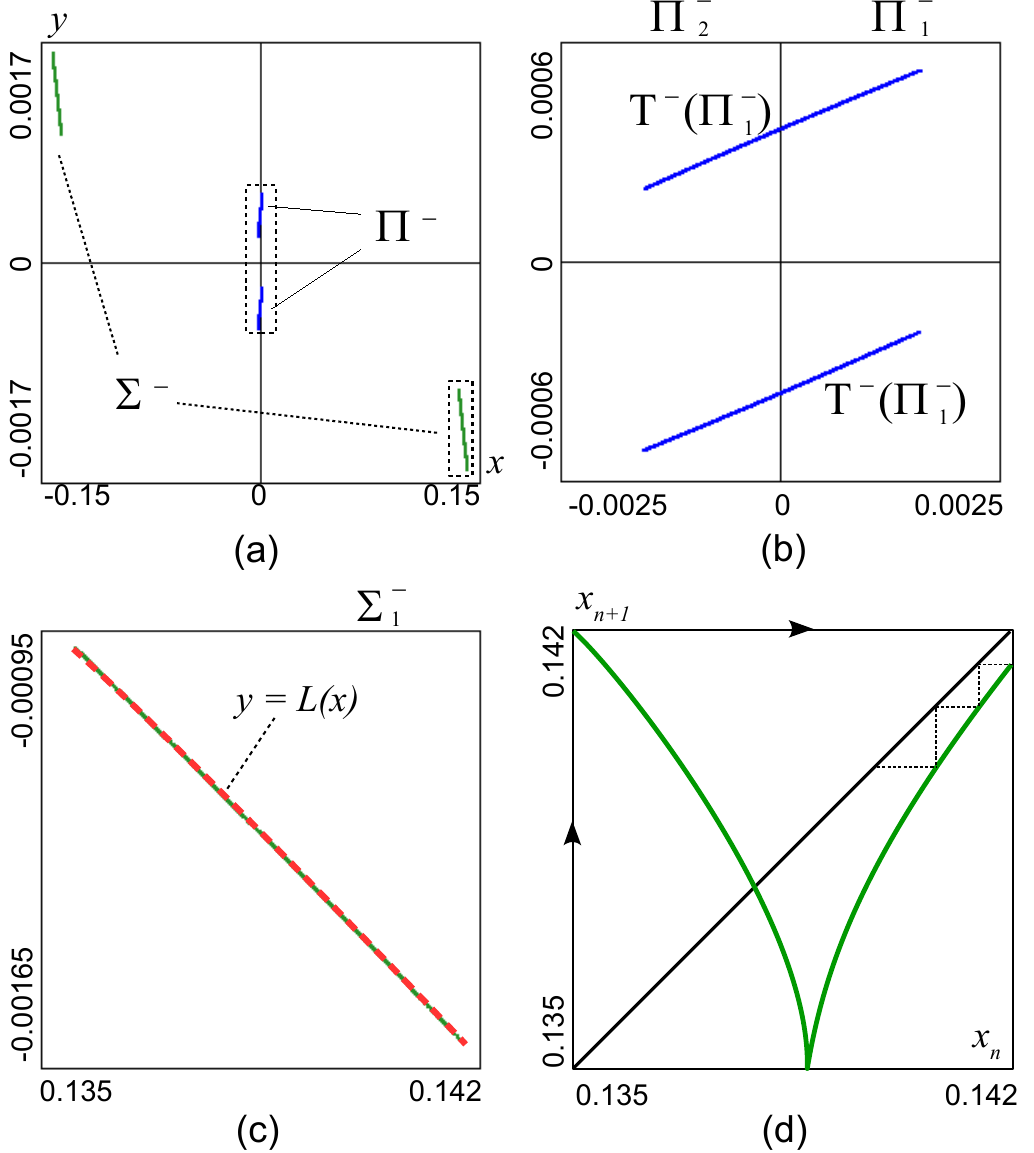} }
\caption{{\footnotesize To an explanation of the procedure for constructing the 2D and 1D first-return maps. As an example we consider the Lorenz attractor of system \eref{eq_mainEq} for $(\alpha, \beta, \mu) = (0.144, 0.15, 0.04)$. (a) We take the unstable separatrix $\Gamma^-_1$ and compute $2 \cdot 10^5$ its intersection points with the cross-section $z=-0.3$, intersections from above are colored in blue (these points lie in the region $\Pi^-$), from below -- in green (these points lie in the region $\Sigma^-$), see the scheme in Fig.~\ref{fig_GM}; (b) enlarged fragment of the 2D first-return map with intersections from above; (c) enlarged fragment of the 2D first-return maps with intersections from below for $x > 0$, the red curve is the graph of interpolating polynomial $y=L(x)$; (d) quotient map constructed by the points of the first return to the cross-section $z=-0.3$ from below.}}
\label{fig_1DMapScheme}
\end{figure}

Further, we perform numerical study of the return maps and their quotients, and compare the results with the predictions of the Lorenz attractor theory. Note that in numerical experiments the wedges $T^+(\Pi^+_{1,2})$ are very thin due to the strong contraction along the leaves of the foliation $F^+_{ss}$. Therefore, to compute the quotient maps, we do not need to actually compute the foliation, as was done e.g. in~\cite{Creaser17}. Instead, we use the following procedure, using the fact the attractor on the cross-section is well-approximated by smooth curves, see Fig.~\ref{fig_1DMapScheme}a.

First, we numerically construct the attractor on a cross-section. For that, we take the unstable separatrix $\Gamma^-_1(O^-)$ and compute $2 \cdot 10^5$ its intersection points with the cross-section $z=-0.3$. The separatrix alternatingly intersects the cross-section from below (upwards) and from above (downwards). The points corresponding to the downward intersections lie in the region $\Pi^-$ (the blue points in Fig.~\ref{fig_1DMapScheme}a), the points of the upward intersections lie in the region $\Sigma^-$ (the green points), see the corresponding enlarged fragments in Fig.~\ref{fig_1DMapScheme}b and Fig.~\ref{fig_1DMapScheme}c, respectively. We approximate one of these curves (the rightmost one\footnote{Strictly speaking, for conformity with the geometric Lorenz model, we need to take the curves lying in $\Pi^-$ as the attractor of the cross-section. However, computationally it is more convenient to take the curves in $\Sigma^-$: the resulting 1D map is the same (up to a coordinate change) because the region $\Sigma^-$ is the image of $\Pi^-$ by the orbits of the system. Also, by the symmetry, it is enough to consider only the curve in $\Sigma^-_1$, see Fig.~\ref{fig_GM}a}) as the graph of an interpolating polynomial $y = L(x)$ (the red curve in Fig.~\ref{fig_1DMapScheme}c). For uniformly distributed values $x_i > 0$, for each point $(x_i, y_i=L(x_i))$ we find its image $(\bar x_i, \bar y_i)$ by the 2D return map $T^+$, and take the map $x_i \mapsto \bar x_i$ as the numeric approximation to the quotient map $\tau$. Utilizing the symmetry, we consider only positive $x$, and study the map $x \mapsto |\tau(x)|$. The graph of the resulting 1D map is shown in Fig.~\ref{fig_1DMapScheme}d. More details on the computation of the quotient maps can be found in~\cite{Kaz2021}.

For the Simo angels we apply the same procedure to compute quotients for both 2D first- and half-return maps $T$ and $\hat T^+$ . The detailed analysis of 2D and 1D maps for the Lorenz attractors and Simo angels is presented in Sections~\ref{sec3_1} and~\ref{sec3_2}, respectively.

\section{Pseudohyperbolic attractors in the normal form} \label{sec3}

\begin{figure}[tb]
\center{\includegraphics[width=1.0\linewidth]{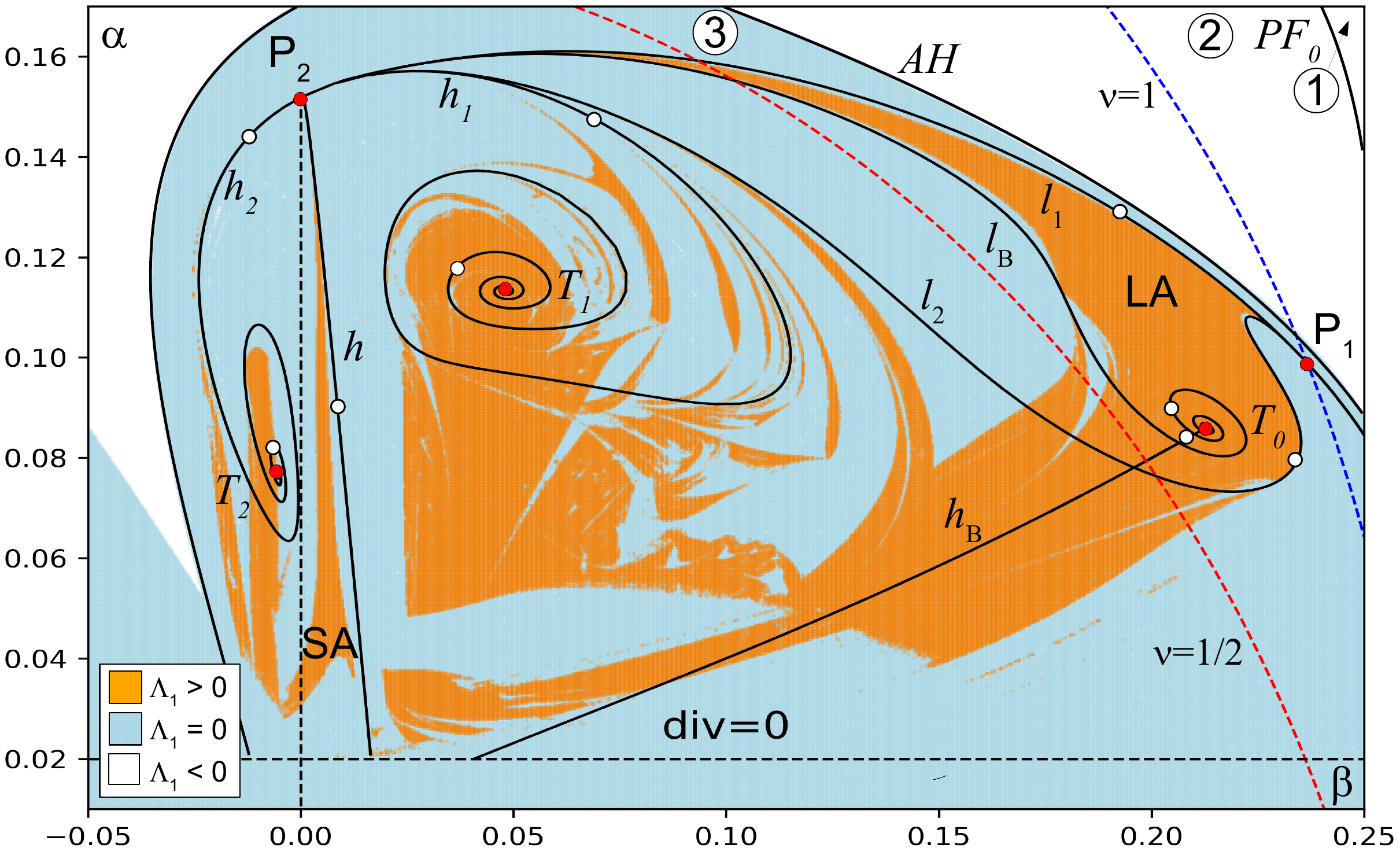} }
\caption{{\footnotesize Colored diagram (see the palette in the bottom-left corner) for the normal form \eref{eq_mainEq} on the parameter plane $(\beta,\alpha)$ for $\mu=0.04$: $PF_0$ -- supercritical pitchfork bifurcation of the equilibria $O^{+}$ and $O^{-}$, stable equilibria $O^+_1$, $O^+_2$ and $O^-_1$, $O^-_2$ are born near $O^{+}$ and $O^{-}$ below this curve; $AH$ -- supercritical Andronov-Hopf bifurcation of the equilibria $O^{\pm}_{1,2}$; $l_1, l_2$, and $l_{B}$ -- homoclinic, doubled homoclinic, and winded onto the Bykov point $T_0$ homoclinic butterflies, see Figs.~\ref{fig2}a--c; $h_B$ -- homoclinic bifurcation of the saddle-focus equilibria $O^\pm_{1,2}$ (Fig.~\ref{fig3}d); curve $h_1 \cup h_2$ separated by the point $P_2$ (where the equilibrium $O$ has a pair of equal negative eigenvalues) onto two parts corresponds to the heteroclinic bifurcation when the unstable separatrices $\Gamma^\pm_{1,2}(O^{\pm})$ return to $O$ (Fig.~\ref{fig3}e,g,h); $h$ -- heteroclinic bifurcation when $\Gamma^\pm_{1,2}(O^{\pm})$ tends to $O^{\mp}$ (Fig.~\ref{fig3}f). Point $P_1$ of the intersection of $l_1$ with the neutral saddle curve $\nu=1$ gives the origin of the Lorenz attractor existence region LA. Point $P_2$ gives the origin for the LA region and for the Simo angel existence region SA. Codimension two points $T_1$ and $T_2$ correspond to Bykov-like cycles, when $\Gamma^\pm_{1,2}(O^{\pm})$ become get to two-dimensional manifolds $W^s(O^{\mp}_{1,2})$. By the circles ~\raisebox{.5pt}{\textcircled{\raisebox{-.9pt} {1}}}, ~\raisebox{.5pt}{\textcircled{\raisebox{-.9pt} {2}}}, and ~\raisebox{.5pt}{\textcircled{\raisebox{-.9pt} {3}}} we mark points for which phase portraits are shown in Fig.~\ref{fig2}, phase portraits for other circles are shown in Fig.~\ref{fig3}.}}
\label{fig1}
\end{figure}

In this Section we present a detailed bifurcation analysis for system \eref{eq_mainEq}. We pay special attention to bifurcations leading to the appearance of pseudohyperbolic attractors. Bifurcation curves on the $(\beta,\alpha)$-parameter plane superimposed with the chart of Lyapunov exponents are presented in Figure~\ref{fig1}. Most of bifurcation curves are found with the help of the MatCont package~\cite{dhooge2008new, de2012interactive}. To compute Lyapunov diagrams we take, at each pair of parameter values, a sufficiently long orbit with initial conditions near the equilibrium $O^+$ and estimate the maximal Lyapunov exponent $\Lambda_1$. Depending on its sign, the corresponding pixel of the $(\beta,\alpha)$-parameter plane is colored according to the palette presented in the bottom-left corner of Fig.~\ref{fig1}.

\begin{figure}[h]
\center{\includegraphics[width=1.0\linewidth]{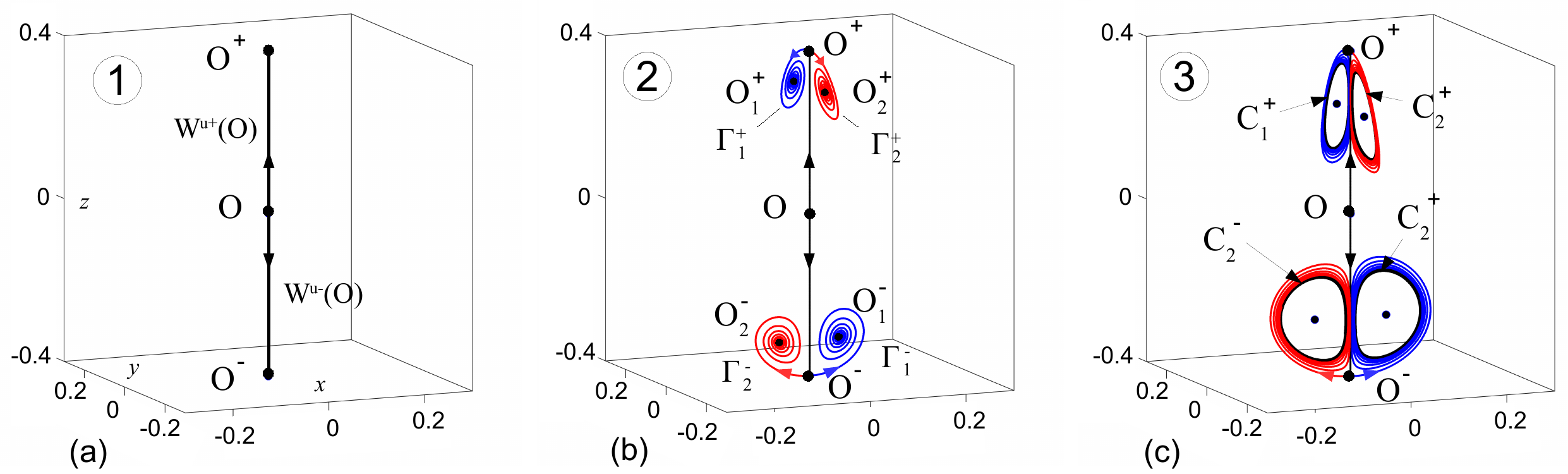} }
\caption{{\footnotesize Phase portraits at the points ~\raisebox{.5pt}{\textcircled{\raisebox{-.9pt} {1}}}, ~\raisebox{.5pt}{\textcircled{\raisebox{-.9pt} {2}}}, and ~\raisebox{.5pt}{\textcircled{\raisebox{-.9pt} {3}}} marked in the bifurcation diagram presented in Fig.~\ref{fig1}.}}
\label{fig2}
\end{figure}

Recall that in system \eref{eq_mainEq} the trivial equilibrium $O(0,0,0)$ is always saddle with a two-dimensional stable and a one-dimensional unstable invariant manifolds $W^s(O)$ and $W^u(O)$. The unstable separatrices $\Gamma^+$ and $\Gamma^-$ forming together with $O$ the unstable manifold $W^u(O)$ always tend to the equilibria $O^{+}$ and $O^{-}$, respectively, see Fig.~\ref{fig2}a, which, depending on parameter values, can be either stable or saddle. Passing through a supercritical pitchfork bifurcation curve $PF_0$, the equilibrium $O^{+}$ (and, by the $\mathcal{S}$-symmetry, $O^{-}$) changes stability. As a result, a pair of stable equilibria $O^+_1$ and $O^+_2$ ($O^-_1$ and $O^-_2$) are born near $O^{+}$ ($O^{-}$) below the $PF_0$-curve, see Fig.~\ref{fig2}b. This pair of equilibria undergoes a supercritical Andronov-Hopf bifurcation on a curve $AH$. As a result, stable limit cycles $C^{\pm}_{1,2}$ are born around $O^\pm_{1,2}$, see Fig.~\ref{fig2}c.

Let $\lambda_2, \lambda_1$, and $\gamma$ be the eigenvalues of the linearization of the system at the equilibria $O^{+}$ and $O^{-}$. Then, after the pitchfork bifurcation $PF_0$, we have $\lambda_2 < \lambda_1 < 0 < \gamma$, and one can define the so-called saddle index $\nu(O^{\pm}) = -\lambda_1 / \gamma$. The curves $\nu=1$ and $\nu=1/2$ are presented in Fig.~\ref{fig1}. We say that on the curve $\nu = 1$ the saddle equilibria $O^{\pm}$ are neutral, above this curve they are contracting, and below -- expanding.

\begin{figure}[h]
\center{\includegraphics[width=1.0\linewidth]{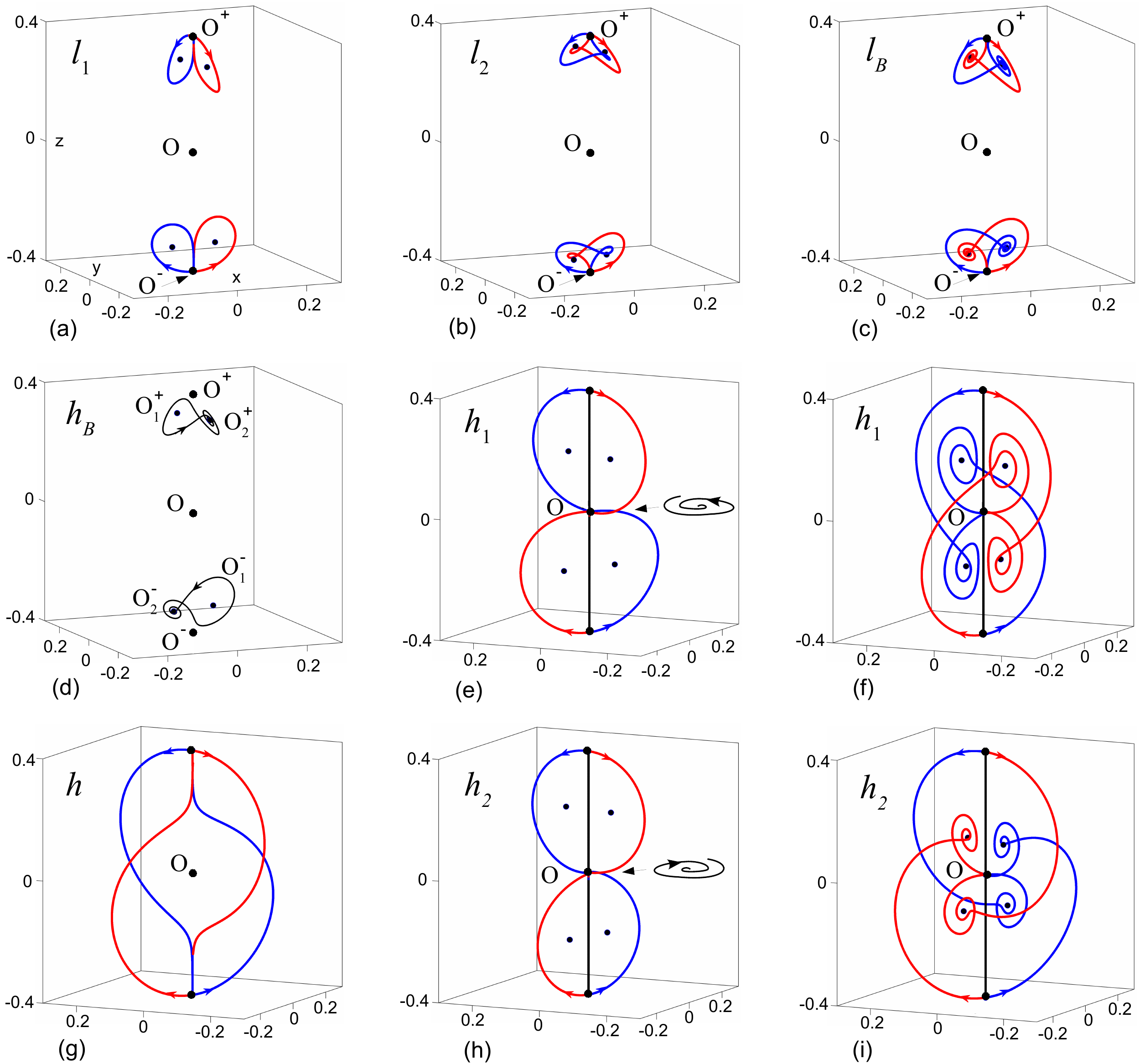} }
\caption{{\footnotesize Phase portraits corresponding to homo- and heteroclinic bifurcations occurring on the curves (a) $l_1$, (b) $l_2$, (c) $l_B$, (d) $h_B$, (e)--(f) $h_1$, (g) $h$, (h)--(i) $h_2$.}}
\label{fig3}
\end{figure}

The curves $l_1, l_2$, and $l_{B}$, shown in Fig.~\ref{fig1}, correspond to homoclinic bifurcations of equilibria $O^{\pm}$. Since system \eref{eq_mainEq} is Z4-symmetric, the appearance of one homoclinic orbit (loop), e.g. for the equilibrium $O^+$, automatically leads to the appearance of a homoclinic butterfly bifurcations (symmetric pairs of homoclinic loops) with both equilibria $O^\pm$, see the corresponding phase portraits in Figs.~\ref{fig3}a,b,c.

The curve $l_B$ winds around the Bykov point $T_0$ corresponding to the appearance of a codimension-two heteroclinic connection between the saddle $O^{+}$ and saddle-focus $O^+_1$. Here, the unstable separatrix $\Gamma^+_1(O^+)$ and one of the stable separatrices of $O^+_1$ coincide, while the two-dimensional manifolds of $O^+$ and $O^+_1$ intersect transversally. Thus, we have here a heteroclinic cycle connecting $O^{+}$ and $O^+_1$. By the Z4-symmetry, we have four such cycles. The phase portrait at the point $B$ is shown in Fig.~\ref{fig3}c. Bifurcations of such heteroclinic cycles were studied in Refs.~\cite{Bykov88, Bykov93}. In particular, it is known from these works that the homoclinic bifurcation curve $h_B$ corresponding to the appearance of a Shilnikov loop of the saddle-focus $O^+_1$ (Fig.~\ref{fig3}d) starts at the point $B$. Note that the curves $l_1, l_2$, and $l_{B}$ play an important role in the structure of the boundary of the region LA of the existence of the Lorenz attractors pair, see more details in Sec.~\ref{sec3_1}.

Curves $h, h_1$, and $h_2$ correspond to heteroclinic bifurcations. On the curves $h_1$ and $h_2$ the unstable separatrices of $O^+$ (and $O^-$, by the symmetry) lie in the two-dimensional stable manifold of the saddle-focus $O$ (Figs.~\ref{fig3}e,f,h,i). On the curve $h$ they tend to the saddle $O^+$ ($O^-$), see Figs.~\ref{fig3}g. Note that, in the restriction to $W^s(O)$, the equilibrium $O$ is a stable focus except for $\beta=0$ where it becomes a dicritical node due to a pair of equal negative eigenvalues. When $\beta>0$ the rotation around $O$ is counterclockwise, when $\beta < 0$ -- clockwise, see Figs.~\ref{fig3}e and~\ref{fig3}h. The region SA of the existence of the Simo angels attractors adjoins to the point P$_2$ where the heteroclinic cycle with the ``dicritical saddle'' $O$ occurs, see Sec.~\ref{sec3_2} for more details.

Among regions with chaotic dynamics the LA- and SA-regions are of special interest, since they are free of visible stability windows. In the following two subsections, we show that, indeed, there are no stability windows in these regions, because the corresponding strange attractors (a pair of the Lorenz attractors, or a two- or four-winged Simo angel) are pseudohyperbolic.

\subsection{The Lorenz attractors in system \eref{eq_mainEq}} \label{sec3_1}

\begin{figure}[tb]
\center{\includegraphics[width=1.0\linewidth]{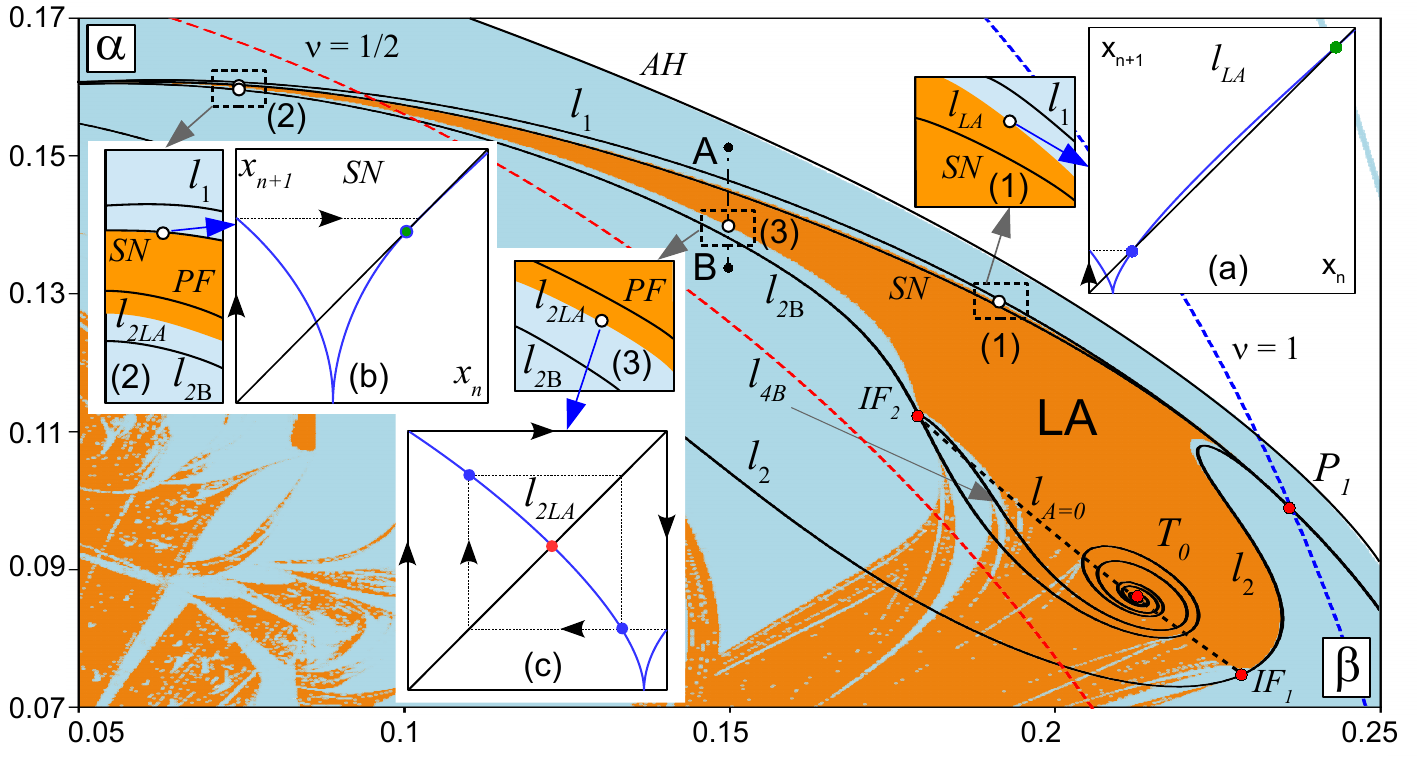} }
\caption{{\footnotesize Organization of the Lorenz attractor existence region LA. $SN$ and $PF$ -- saddle-node and subcritical pitchfork bifurcations of periodic orbits. $l_{LA}, l_{2LA}, l_{4LA}, \dots$ -- heteroclinic bifurcations when the unstable separatrices $\Gamma^+_{1,2}(O^+)$ get the saddle periodic orbits which are born below the curves $l_1, l_{2B}, l_{4B}, \dots$. Illustrations for the bifurcations occurring on the curves $l_{LA}, SN$, and $l_{2LA}$ with the help of the 1D first-return maps are shown in the insets (a), (b), and (c), respectively. The insets (1), (2), and (3) show the enlarged fragments of the diagram. $l_{2B}$ and $l_{4B}$ -- two-round and four-round homoclinic butterflies, these curves are winded onto the Bykov point $T_0$. Codimension-two points IF$_1$ and IF$_2$ on the curves $l_2$ and $l_{2B}$ correspond to the inclination flip bifurcations. Between these curves the boundary of the LA-region is associated with the curve $l_{A=0}$ on which the subspaces $E^{ss}$ and $E^{cu}$ intersect non-transversally. Codimension-two point P$_1$ -- homoclinic butterflies to the neutral saddles $O^+$ and $O^{-}$.}}
\label{fig4}
\end{figure}

The detailed bifurcation diagram for the LA-region is shown in Fig.~\ref{fig4}. We supplement this diagram with enlargements of some of its fragments and with 1D first-return maps (numerically evaluated quotients of the 2D return maps).

For $1/2 < \nu < 1$ the bifurcation diagram is very similar to the bifurcation diagram for the Shimizu-Morioka system~\cite{ASh86, ASh93}. This is not a coincidence. Indeed, as shown in~\cite{KKST24}, a suitable renormalization of system~\eref{eq_mainEq} has, in the limit $\mu=0$, a pair of symmetric equilibrium states $O^+$ and $O^-$ undergoing triple-zero bifurcation for which the Shimizu-Morioka system is a normal form. Therefore, as we take $\mu$ sufficiently small, in this part of the $(\beta,\alpha)$-plane we observe the same classical scenarios of the creation of the Lorenz attractor as in the Shimizu-Morioka system. These scenarios were proposed in~\cite{ABS82, Sh80}.

As an example consider the rectangle (1) in the bifurcation diagram (inset (1) in Fig.~\ref{fig4}). Above the curve $l_1$ the unstable separatrices $\Gamma^+_1$ and $\Gamma^+_2$ tend to the equilibria $O^+_1$ and $O^+_2$, respectively. The curve $l_1$ corresponds to the homoclinic butterfly to $O^+$ (by the symmetry, we also have a homoclinic butterfly to $O^-$). Upon crossing $l_1$, the butterfly splits, and a pair of saddle periodic orbits $C^+_1$ and $C^+_2$ together with a nontrivial hyperbolic set $\Delta^+$ are born. This set becomes the Lorenz attractor below the curve $l_{LA}$ on which the unstable separatrices $\Gamma^+_1$ and $\Gamma^+_2$ get to the stable two-dimensional manifolds of the periodic orbits $C^+_2$ and $C^+_1$, respectively. An illustration of this bifurcation for the numerically computed 1D first-return map is shown in the inset (a) in Fig.~\ref{fig4}. An orbit starting at $x=0$ corresponds to the unstable separatrix $\Gamma^+_2$, the left fixed point corresponds to the saddle periodic orbit $C^+_1$. By the $\mathcal{S}$-symmetry, in system \eref{eq_mainEq} after this bifurcation a pair of the Lorenz attractors is born.

The second part of the upper boundary of the LA-region, to the left of the curve $\nu=1/2$, is a saddle-node bifurcation curve $SN$ where the saddle cycles $C^+_1$ and $C^+_2$ merge with the stable ones, see inset (2). The illustration for this bifurcation for the 1D first-return map is shown in inset (b).

The lower part of the boundary of the LA-region consists of many fragments $l_{2LA}$, $l_{4LA}, \dots$ corresponding to heteroclinic bifurcations of the unstable separatrices $\Gamma^+_{1,2}(O^+)$ lying on the stable manifolds of, respectively, 2-round, 4-round, $\dots$ saddle periodic orbits, the illustration of this for the curve $l_{2LA}$ is shown in insets (3) and (c).

Let us describe in details bifurcations along the pathway AB: $\beta=0.15, \alpha \in [0.1451, 0.1397]$ crossing the upper and lower boundaries of the LA-region at $1/2 < \nu < 1$. The corresponding 2D and 1D first-return maps are shown in Figure~\ref{fig5}. The Lorenz attractor is born below the curve $l_{LA}$ where the separatrices $\Gamma^+_1(O^+)$ and $\Gamma^-_2(O^+)$ lie on the stable manifold of the periodic orbits $C^+_2$ and $C^+_1$, respectively. Here, the newly born Lorenz attractor coexists with a pair of stable fixed points $S^+_{1,2}$, see Fig.~\ref{fig5}a. Then, on the curve $SN$ the stable fixed points $S^+_{1}$ and $S^+_{2}$ merge with saddles $C^+_1$ and $C^+_2$, see Fig.~\ref{fig5}b, and disappear altogether. The next important bifurcation occurs when both unstable separatrices $\Gamma^+_1(O^+)$ and $\Gamma^+_2(O^+)$ get the stable manifold of the symmetric 2-round saddle cycle $C^+_0$, see Fig.~\ref{fig5}c. After this bifurcation a lacuna occurs inside the attractor: the periodic orbits $C^+_0$ is no longer its part, see Fig.~\ref{fig5}d. On the curve $PF$ the periodic orbit $C^+_0$ undergoes the subcritical pitchfork bifurcation: it becomes stable below this line, and a pair of saddle periodic orbits $C^+_{01}$ and $C^+_{02}$ is born, see Fig.~\ref{fig5}e. The Lorenz attractor disappears below the curve $l_{2LA}$ where the unstable separatrices $\Gamma^+_1(O^+)$ and $\Gamma^+_2(O^+)$ lie on the stable manifolds of $C^+_{01}$ and $C^+_{02}$, see Fig.~\ref{fig5}f.

\begin{figure}[h]
\center{\includegraphics[width=1.0\linewidth]{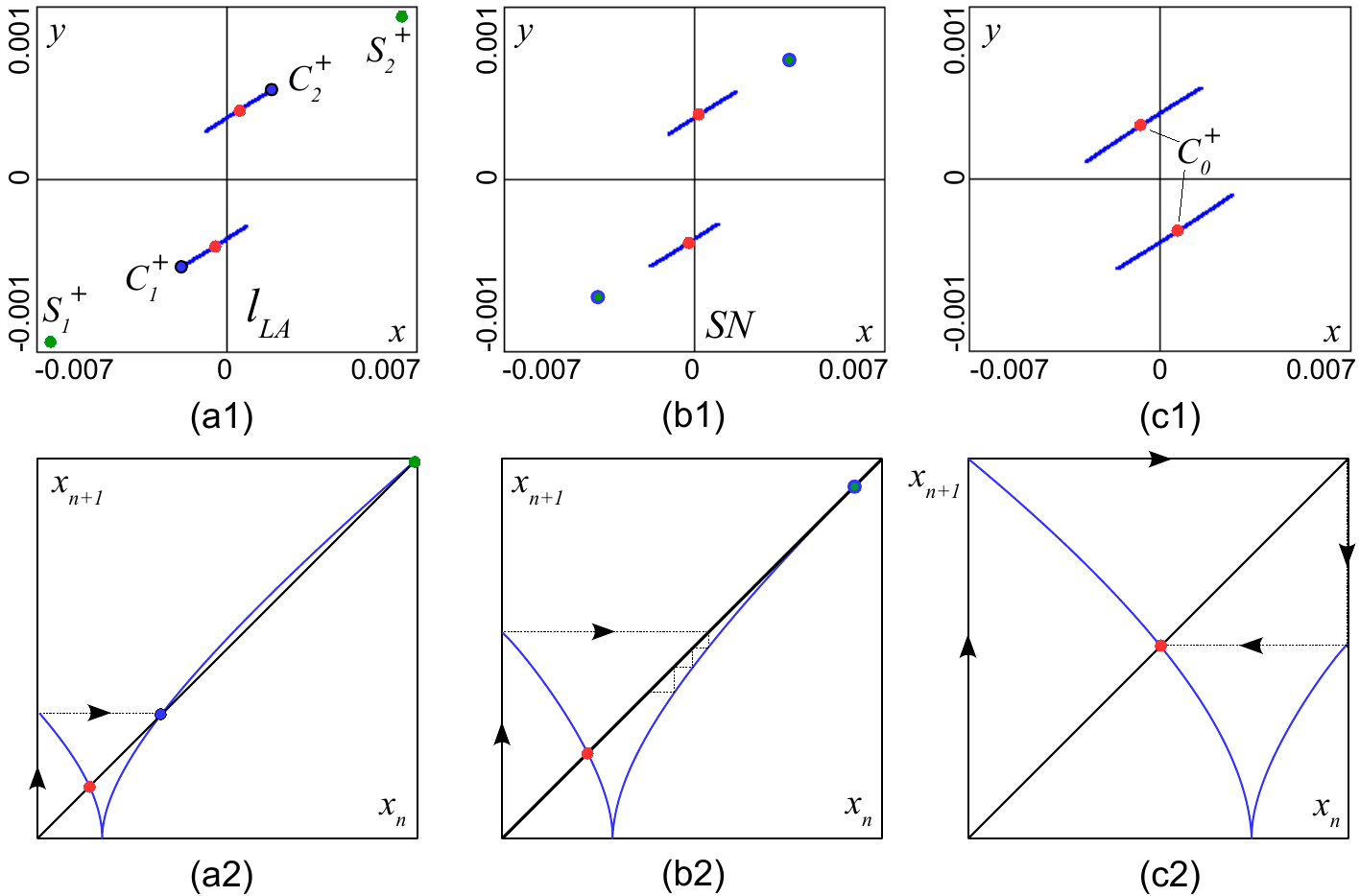} }
\caption{{\footnotesize Bifurcations along the pathway AB: $\beta=0.15, \alpha \in [0.1451, 0.1397]$: first row -- 2D first-return maps with the cross-section $z=0.3$; second row -- corresponding quotient maps. (a) $\alpha=0.1451$, (b) $\alpha=0.1449$, (c) $\alpha=0.1426$, (d) $\alpha=0.142$, (e) $\alpha=0.14$, (f) $\alpha=0.1397$.}}
\label{fig5}
\end{figure}

\setcounter{figure}{7}

\begin{figure}[h]
\center{\includegraphics[width=1.0\linewidth]{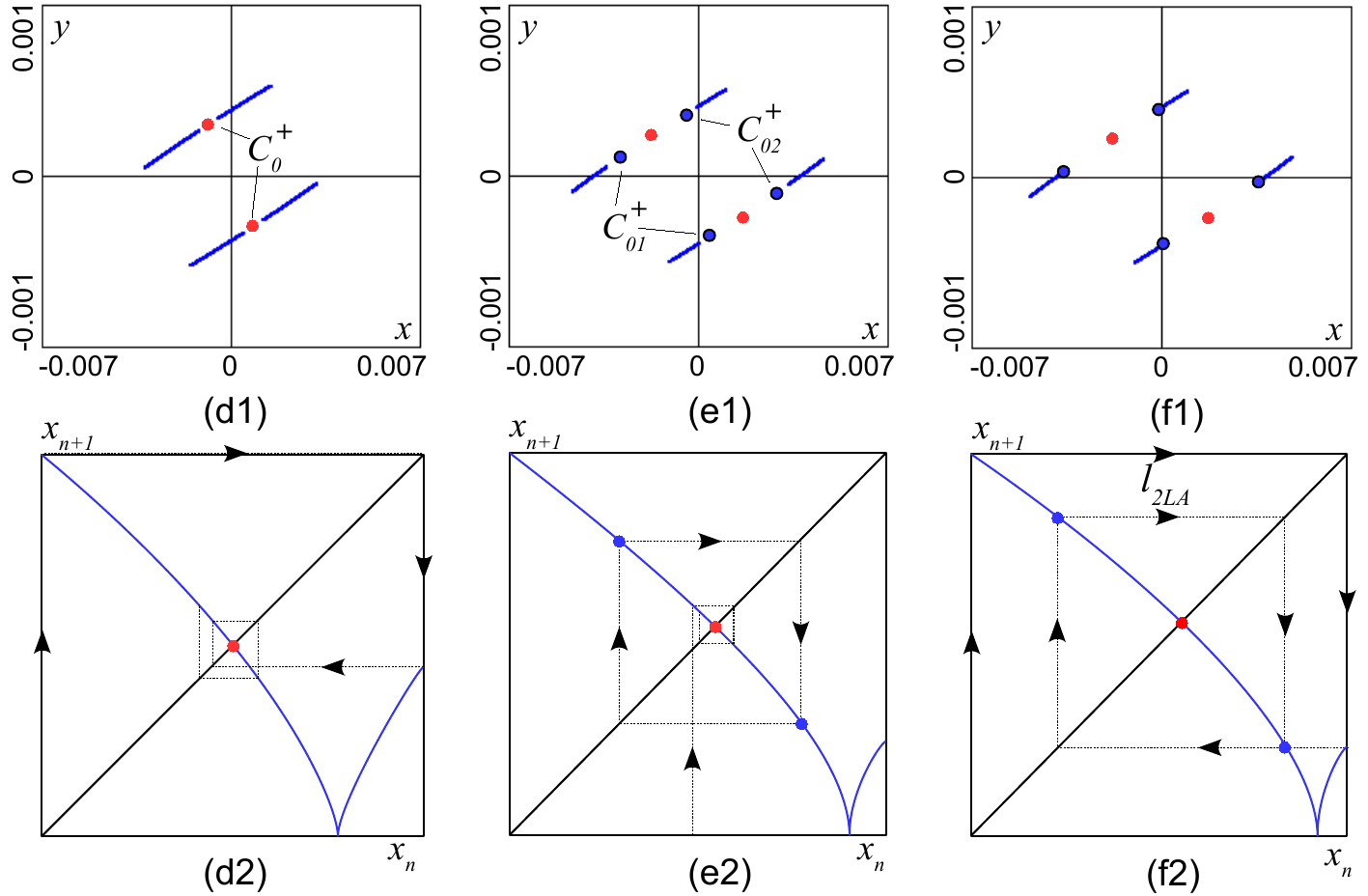} }
\caption{{\footnotesize (Continued).}}
\end{figure}

The right boundary of the LA-region is associated with the curve $l_2$. To the left of this curve, a pair of doubled saddle cycles $C^{2+}_1$ and $C^{2+}_2$ together with a nontrivial hyperbolic set are born from the corresponding doubled homoclinic butterfly. This hyperbolic set become the Lorenz attractor when $\Gamma^+_1(O^+)$ and $\Gamma^+_2(O^+)$ get on the stable manifolds of $C^{2+}_1$ and $C^{2+}_2$, respectively\footnote{This heteroclinic bifurcation curve is very close to the curve $l_2$, thus, it is not shown in Fig.~\ref{fig4}.}.

\begin{figure}[h]
\center{\includegraphics[width=0.7\linewidth]{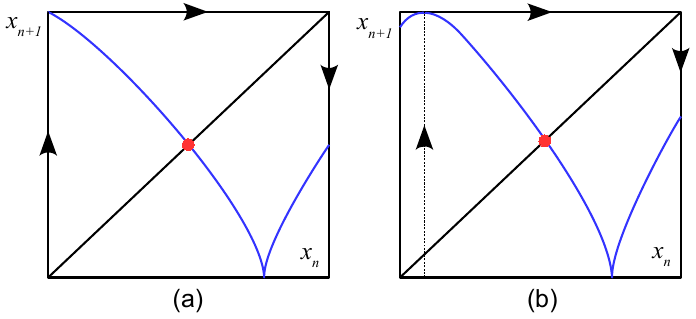} }
\caption{{\footnotesize 1D first-return maps above and below the curve $l_{A=0}$. $\beta=0.2$: (a) $\alpha=0.1$, (b) $\alpha=0.092$.}}
\label{fig6}
\end{figure}

The bottom boundary of the LA-region region is most complicated, it is associated with the curve $l_{A=0}$, where the strong stable and central unstable subspaces $E^{ss}$ and $E^{cu}$ intersect non-transversely. For the Lorenz system such a curve was found by Bykov and A.~Shilnikov in \cite{BASh92} (see also \cite{AShil2012, Creaser17}) by means of the 1D first-return maps analysis. For the Shimizu-Morioka system this curve was computed by A.~Shilnikov in \cite{ASh86, ASh93} Above the curve $l_{A=0}$, the corresponding 1D map has non-zero derivative, see e.g. Fig.~\ref{fig6}a. Below this curve, a critical point with zero derivative appears which manifests itself in characteristic hook-shape of the attractor on the cross-section, see Fig.~\ref{fig6}b.

In system \eref{eq_mainEq}, we have $1/2 < \nu < 1$ along the curve $l_{A=0}$. Therefore, this curve does not form the exact boundary of the LA-region~\cite{ASh93} (this is different from the Lorenz system where $0 < \nu < 1/2$ along $l_{A=0}$). The fact that $\nu > 1/2$ implies that stable periodic orbits exist above this curve, and the Lorenz attractor is formed after additional heteroclinic bifurcations at the curves $l_{2LA}$, $l_{4LA}, \dots$ mentioned above; a detailed description of the fractal boundary of the LA-region above the curve $l_{A=0}$ see in \cite{ASh93}.

Below the curve $l_{A=0}$, we observe the so-called \textit{Shilnikov flames} -- the region of existence of non-orientable Lorenz attractors~\cite{ASh93, SST93, AShil2014}. It is important to note that the non-orientable Lorenz attractors are pseudohyperbolic (for the Shimizu-Morioka model the pseudohyperbolicity of these attractors was confirmed in \cite{GKKT21}). Thus, robustly chaotic attractors exist also below $l_{A=0}$.

As we mentioned, the structure of the LA-region in system \eref{eq_mainEq} is very similar to that in the Shimizu-Morioka system (cf. Fig.~\ref{fig1} with e.g. Fig.~1 in \cite{ASh93}). In both systems the LA-region adjoins the codimension-two points $P_1$ corresponding to the homoclinic butterfly to a neutral saddle equilibrium and $IF_1$ where the separatrix values for doubled homoclinic loops (occurring on the curve $l_2$) vanish. By the Shilnikov criterion \cite{Sh81}, a region with the Lorenz attractor indeed originates from these codimension-2 points. However, in system \eref{eq_mainEq} there exists a third codimension-2 point belonging to the boundary of the LA-region. It is the point P$_2$ where a heteroclinic connection between $O^+$ and the saddle equilibrium $O$ with a pair of negative multiple eigenvalues occurs. In the next section, we study how the SA-region with pseudohyperbolic attractors also adjoins to this point.

\subsection{Simo angels in system \eref{eq_mainEq}}  \label{sec3_2}

\begin{figure}[tb]
\center{\includegraphics[width=0.9\linewidth]{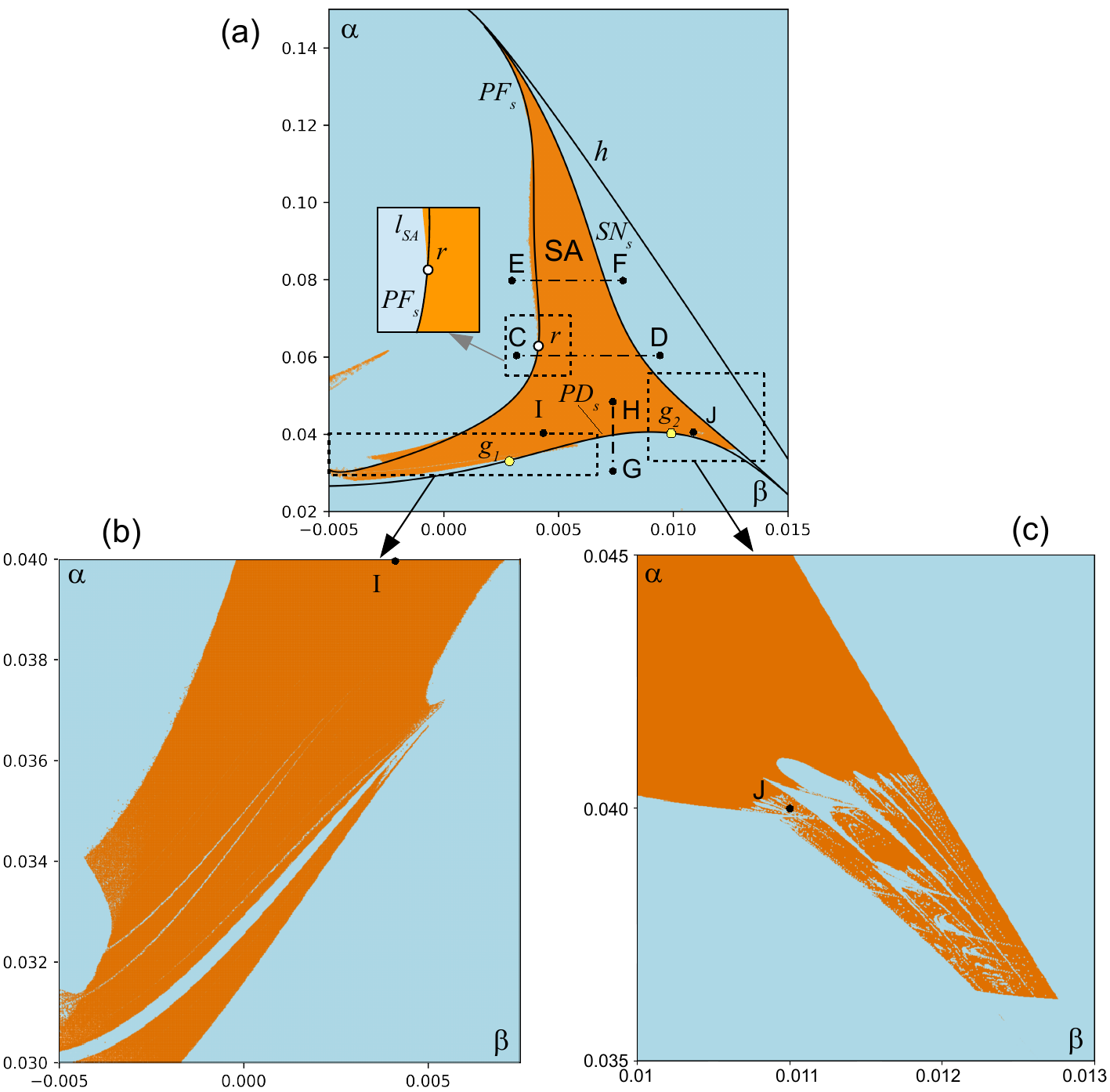} }
\vspace{-0.3cm}
\caption{{\footnotesize (a) Organization of the Simo angels existence region SA. $h$ -- heteroclinic bifurcation when the unstable separatrices $\Gamma^\pm_{1,2}(O^{\pm})$ get to the stable manifolds $W^s(O^{\mp})$, a pair of periodic saddle orbits are born as a result of this heteroclinic bifurcation to the left of the curve $h$; these two periodic orbits merge with the stable ones on the saddle-node bifurcation curve $SN_s$; $l_{SA}$ -- heteroclinic bifurcation when $\Gamma^\pm_{1,2}(O^{\pm})$ get to the stable manifolds of the pair of periodic saddle orbits; $PF_s$ -- subcritical pitchfork bifurcation of a symmetric periodic orbit; $PD_s$ -- period-doubling bifurcation: this bifurcation is subcritical between the points $g_1$ and $g_2$ corresponding to a degenerate period-doubling bifurcation. CD, EF, and GH three pathways along which we study bifurcations leading to the appearance and destruction of the Simo angels. (b) and (c) two enlarged fragments of the Lyapunov diagram marked by the rectangles in the panel (a).}}
\label{fig7}
\end{figure}

\begin{figure}[h]
\center{\includegraphics[width=1\linewidth]{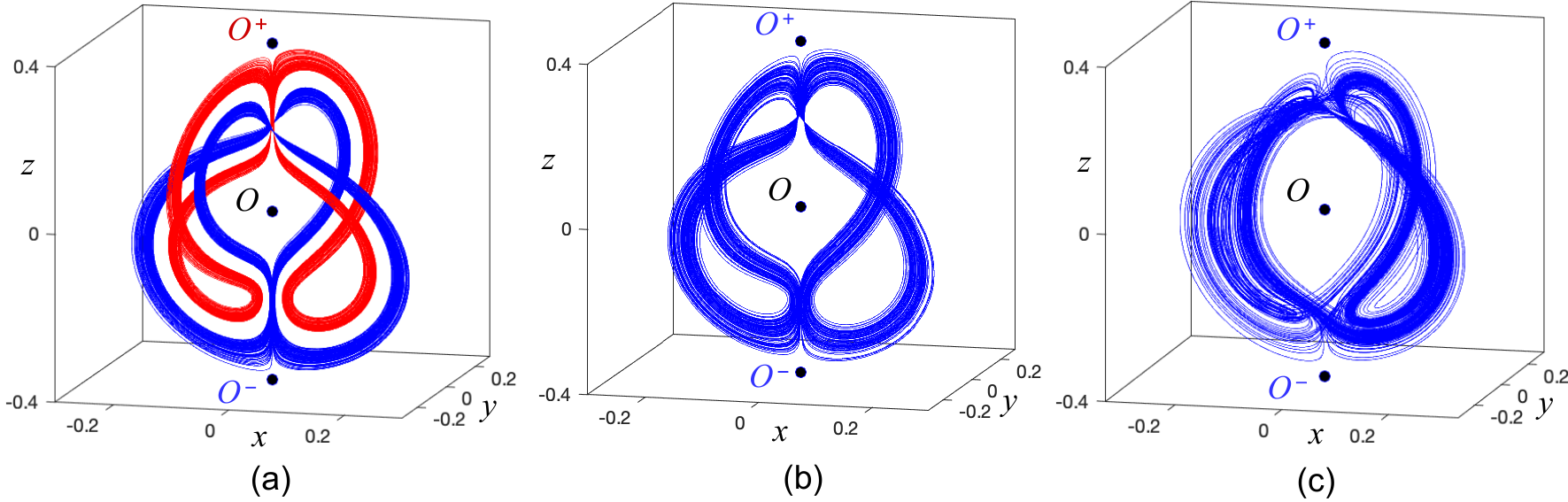} }
\caption{{\footnotesize Phase portraits for different types of Simo angels: (a) $(\alpha, \beta) = (0.08, 0.006)$, four-winged attractor containing both equilibria $O^+$ and $O^-$; (b) $(\alpha, \beta) = (0.08, 0.004)$, a pair of two-winged attractors, one of them contains $O^+$, another one -- $O^-$; (c) $(\alpha, \beta) = (0.04, 0.0075)$, four-winged attractor containing both equilibria $O^+$ and $O^-$ above the curve $PD_s$.}}
\label{fig8}
\end{figure}

The region SA in the $(\beta,\alpha)$ plane corresponding to the existence of Simo angels is shown in Figure~\ref{fig7} (this is an enlarged fragment of Fig.~\ref{fig1}). The right boundary of this region is formed by a curve $SN_s$ on which a saddle-node bifurcation of a symmetric periodic orbit occurs. To the right of this curve a pair of symmetric periodic orbits (stable and saddle) exists, the stable orbit attracts almost all orbits from the neighborhoods of $O^+$ and $O^-$. To the left of this curve the four-winged Simo angel appears, see Fig.~\ref{fig8}a.

The left boundary of the SA-region consists of two fragments. Above the point $r$ the boundary is a heteroclinic bifurcation curve $l_{SA}$ for which the unstable separartices $\Gamma^+_{1,2}(O^+)$ get to the stable manifold of a non-symmetric two-round saddle periodic orbit $C_1$. By the symmetry \eref{eq_Z4sym}, the unstable separatrices of $O^-$ lie on the stable manifold of another saddle periodic orbit $C_2$ which is $\mathcal{S}$-symmetric to $C_1$. A pair of two-winged Simo angels is born to the right of $l_{SA}$, see Fig.~\ref{fig8}a. One of them contains $O^+$, another -- $O^-$.

Below $r$, the SA-region is bounded by the subcritical pitchfork bifurcation curve $PF_s$. Here, to the left of $PF_s$, a symmetric two-round stable periodic orbit $C_0$ exists, which attracts orbits from the neighborhoods of $O^+$ and $O^-$. The pair of saddle periodic orbits $C_1$ and $C_2$ merges with the stable periodic orbit $C_0$ on the curve $PF_s$, and the four-winged angel containing both equilibria $O^+$ and $O^-$ is born to the right of $PF_s$. This attractor is shown in Fig.~\ref{fig8}b.

The bottom boundary of the SA-region also consists of several fragments. Between the points $g_1$ and $g_2$, it is formed by a subcritical period-doubling bifurcation occurring on a curve $PD_s$. Below this curve, four stable periodic orbits exist, which attract orbits from the neighborhood of equilibria $O^+$ and $O^-$ (each of the four separatrices $\Gamma^{\pm}_{1,2}(O^{\pm})$ is attracted to its own stable periodic orbit). Above this curve, these periodic orbits become saddle due to the collision with saddle double-round periodic orbits, and the four-winged Simo angel is born, see Fig.~\ref{fig8}c. To the left of the point $g_1$, as well as to the right of $g_2$, the period-doubling bifurcation is supercritical. The transition to the Simo angel attractor here, as well as near the bottom part of the curve $PF_s$, is more complicated and we do not study it in this paper.

The curves $PF_s, SN_s$ and $h$ were found with the help of MatCont package~ \cite{dhooge2008new, de2012interactive}. All three curves end at the point P$_2$ on the line $\beta=0$. At this point the unstable separatrices $\Gamma^\pm_{1,2}$ lie in the two-dimensional stable manifold of the point $O$ which has two equal real negative eigenvalues ($-\alpha$). This is in agreement with the theory of Ref.~\cite{KKST24}.

Recall that the orange color in Fig.~\ref{fig7} corresponds to the positivity of the numerically computed maximal Lyapunov exponent. As we mentioned, this is not enough to claim that the attractor is indeed chaotic: one also needs to check the pseudohyperbolicity. Not in all of the orange region of Fig.~\ref{fig7}a the chaotic attractors are pseudohyperbolic. The stability windows at the bottom part of the SA-region confirm this fact, see the enlarged fragments of the Lyapunov diagram in Figs.~\ref{fig7}b,c. In the upper part of the SA-region the pseudohyperbolicity is restored, see Sec.~\ref{sec3_3}. Theoretically the pseudohyperbolicity of the Simo angel near the point P$_2$ is shown in \cite{KKST24}.

\subsubsection{Scenarios of the angels appearance.}

\begin{figure}[tb]
\center{\includegraphics[width=1.0\linewidth]{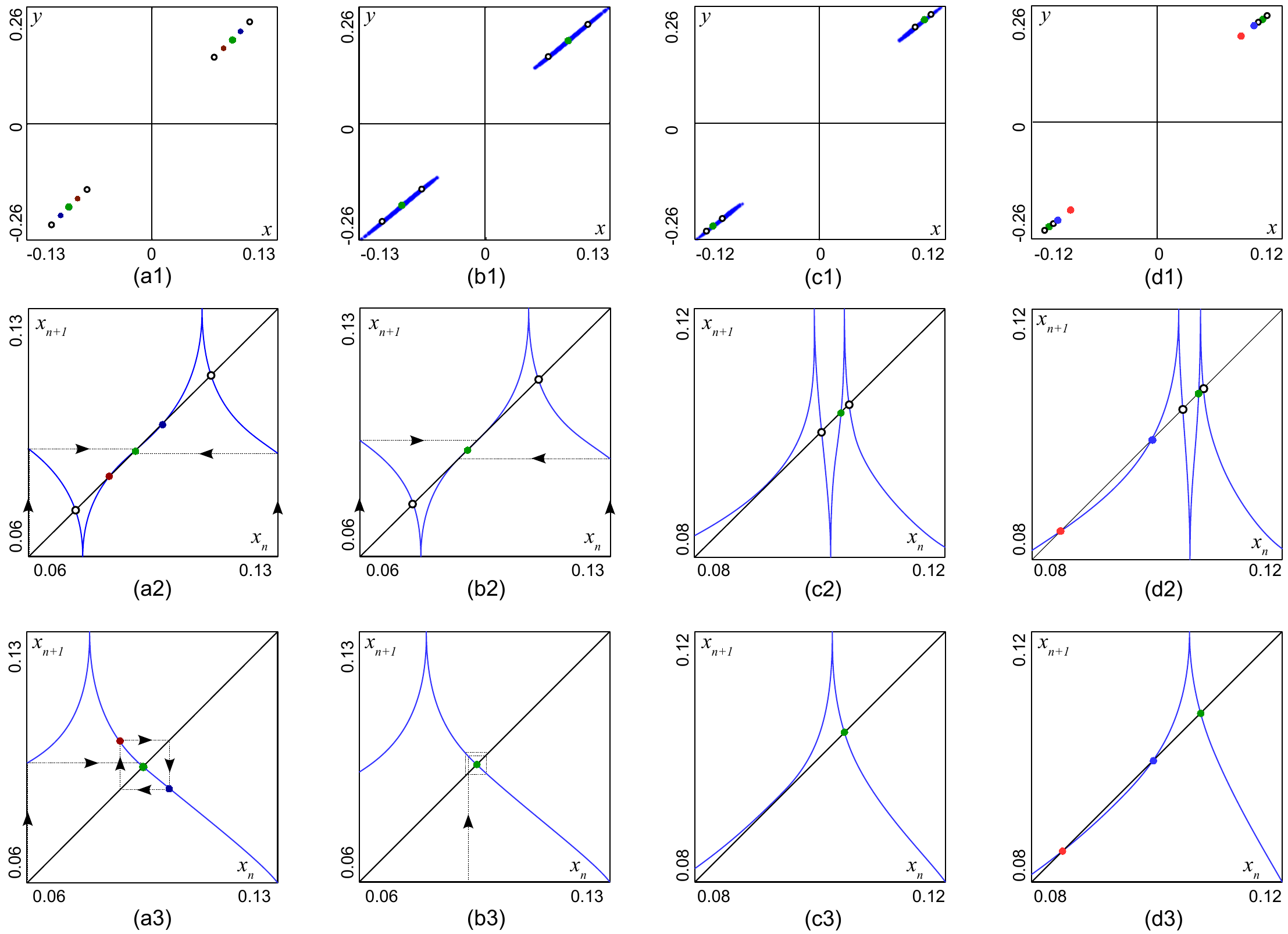} }
\caption{{\footnotesize Bifurcations along the pathway CD: $\alpha=0.06, \beta \in [0.0038, 0.0088]$: first row -- 2D first-return maps $T$ of the cross-section $z=0$; second row -- corresponding quotient maps; third row -- quotients of the 2D half-return maps $\hat T^+$ from the cross-section $\Pi^+: z = 0.1$ to the cross-section $\Pi^-: z = -0.1$. (a) $\beta=0.0038$, (b) $\alpha=0.0041$, (c) $\alpha=0.0086$, (d) $\alpha=0.0087$.}}
\label{fig9}
\end{figure}

A detailed picture of the transformations of the Simo angel is described below with the help of the 2D and 1D return maps introduced in Sec.~\ref{sec2}. We start with the pathway CD: $\alpha=0.06, \beta \in [0.0038, 0.0088]$ corresponding to the formation of the four-winged Simo angel shown in Fig.~\ref{fig8}b. The corresponding illustrations are given in Figure~\ref{fig9}. The top row corresponds to the 2D first-return maps $T$ of the cross-section $z=0$ to itself, the middle row -- quotients of these 2D maps, and in the bottom row we plot quotients of the 2D half-return maps $\hat T^+$ from the cross-section $z = 0.1$ to the cross-section $z = -0.1$.

To the left of the curve $PF_s$ below the point $r$ a symmetric fixed point (bold green) corresponding to the periodic orbit $C_0$ attracts almost all orbits from neighborhoods of the equilibria $O^+$ and $O^-$. A pair of non-symmetric fixed points (bold blue and brown) corresponding to the saddle periodic orbits $C_1$ and $C_2$ exists near it. For the half-return map $\hat T^+$ these non-symmetric points correspond to a period-2 orbit. Illustrations for this case are shown in Fig.~\ref{fig9}a. On the curve $PF_s$, the non-symmetric points merges with the stable fixed point and the four-winged attractor is born to the right of this curve, see Fig.~\ref{fig9}b. Note that this attractor contains both equilibria $O^+$ and $O^-$, the attractor is shown in Fig.~\ref{fig8}b.

At the curve $SN_s$ (see Fig.~\ref{fig9}c), the attractor gets destroyed due to the saddle-node bifurcation. After this a pair of symmetric fixed points appears near the former attractor. The stable fixed point (colored in red) attracts almost all orbits from the neighborhoods of $O^+$ and $O^-$, see Fig.~\ref{fig9}d.

\begin{figure}[b]
\center{\includegraphics[width=1.0\linewidth]{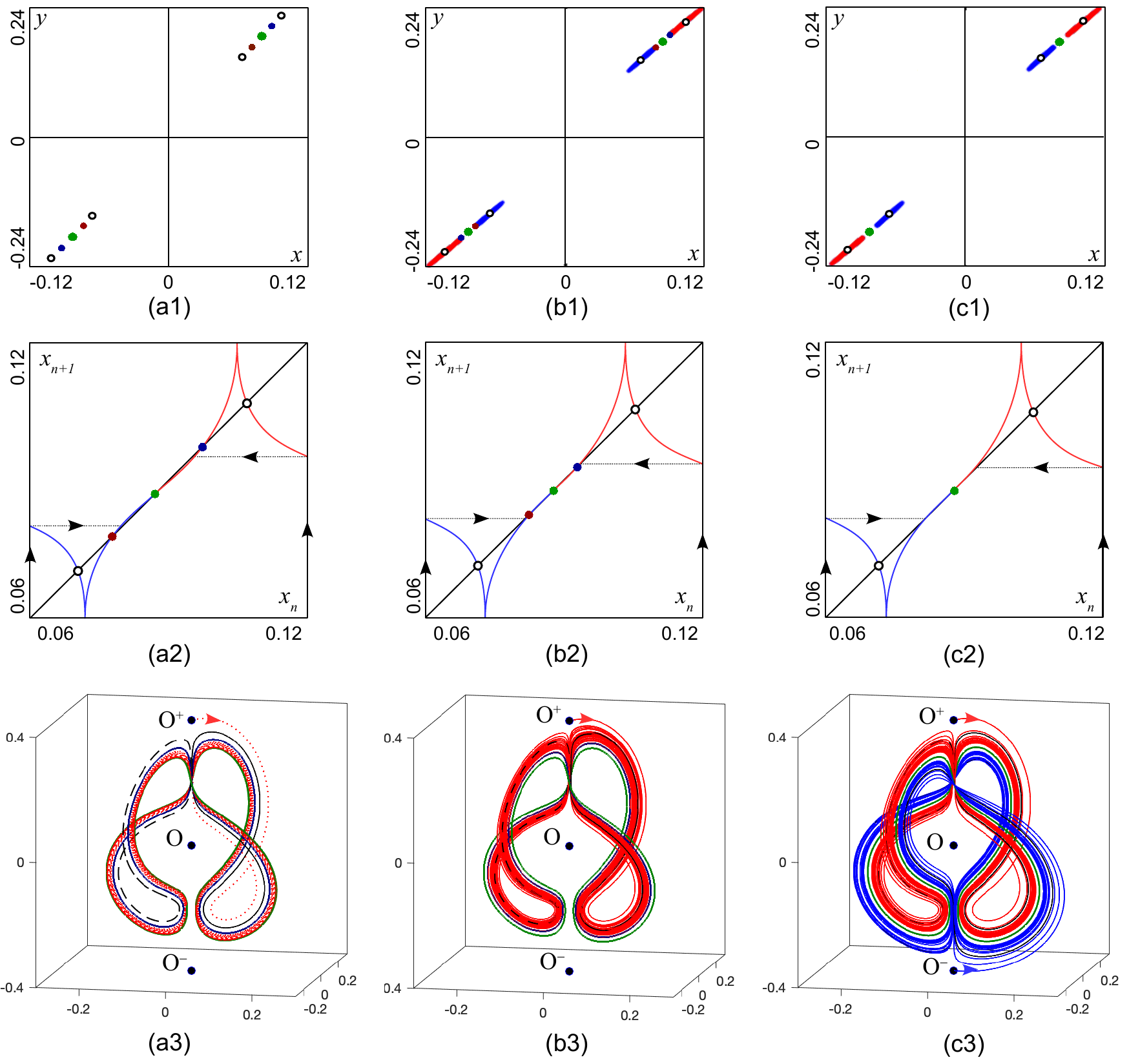} }
\caption{{\footnotesize Bifurcations along the pathway EF: $\alpha=0.08, \beta \in [0.0038, 0.0072]$: first row -- 2D first-return maps with the cross-section $z=0$; second row -- quotients of these maps; third row -- corresponding phase portraits. (a) $\beta=0.0038$, (b) $\beta=0.00396$, (c) $\beta = 0.00405$, (d) $\beta = 0.00475$, (e) $\beta = 0.006$, (f) $\beta = 0.0072$.}}
\label{fig10}
\end{figure}

\begin{remark}
Note that the transition ``a symmetric fixed point $\to$ a chaotic attractor $\to$ a symmetric fixed point'' is different from the transition through the LA-region (compare Figs.~\ref{fig4}b,c and Figs.~\ref{fig9}c3, b3). Such transition is theoretically described in~\cite{SafTur25}.
\label{rem1}
\end{remark}

\setcounter{figure}{12}

\begin{figure}[b]
\center{\includegraphics[width=1.0\linewidth]{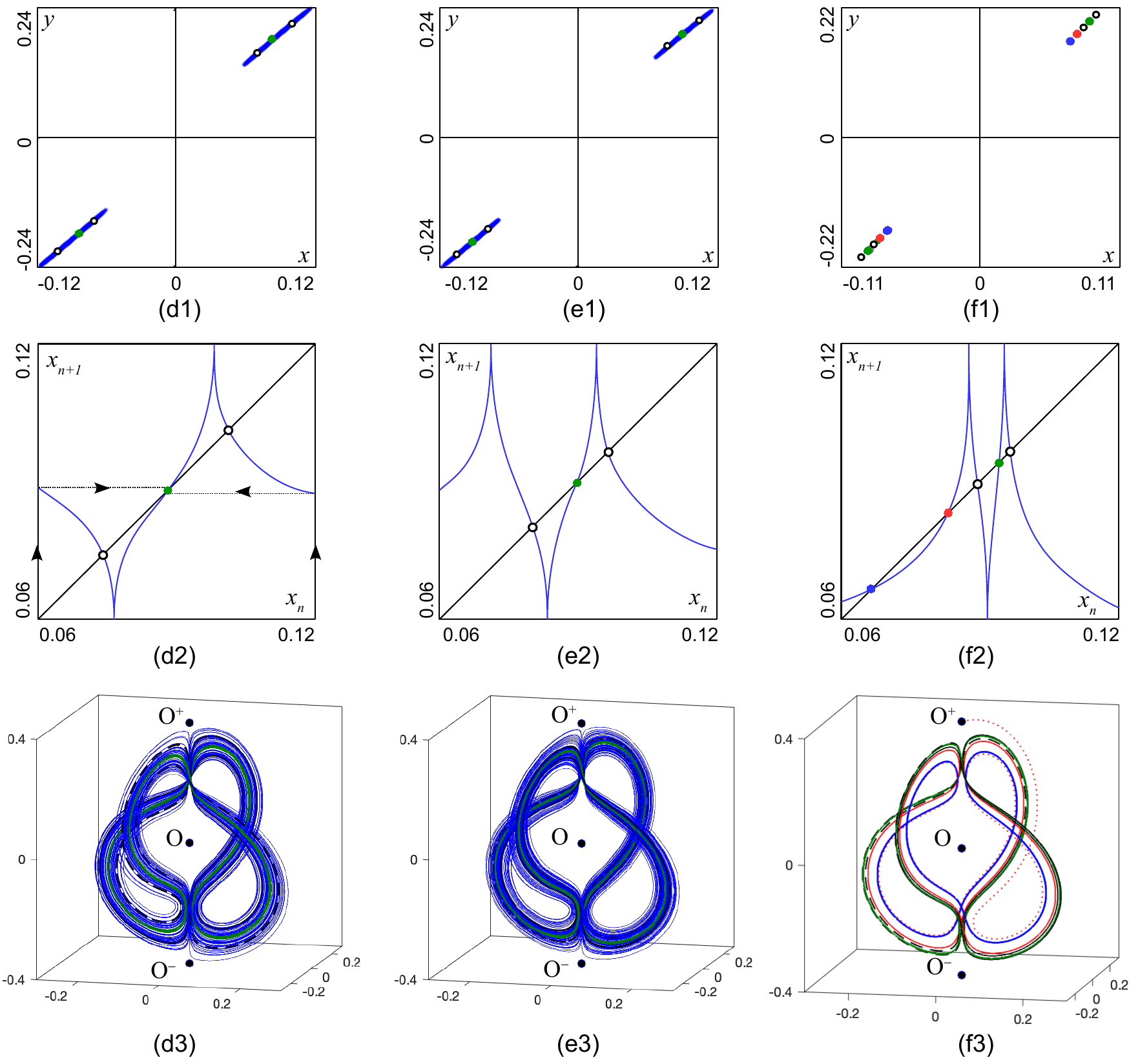} }
\caption{{\footnotesize (Continued).}}
\end{figure}

Another scenario leading to the birth of the Simo angel corresponds to the pathway EF: $\alpha=0.08, \beta \in [0.0038, 0.0072]$ which lies above the point $r$, see Figure~\ref{fig10}. The first row corresponds to 2D first-return maps $T$ of the cross-section $z=0$, the second row -- the quotients of these maps, and the third row -- corresponding phase portraits. As in the previous case, at the beginning, the symmetric stable fixed point (corresponding to the periodic orbit $C_0$) attracts orbits from the neighborhoods of $O^+$ and $O^-$, see Fig.~\ref{fig10}a. In Fig.~\ref{fig10}a3, this orbit is colored in green, a pair of periodic orbits $C_1$ and $C_2$ are colored in blue and brown. Then, a heteroclinic bifurcation occurs: the unstable separatrices $\Gamma^+_{1,2}(O^+)$ get to the stable manifold of $C_1$. By the symmetry, $\Gamma^-_{1,2}(O^-)$ lie on the stable manifold of $C_2$. As a result, the pair of two-winged Simo angels are born after this heteroclinic bifurcation, see Fig.~\ref{fig10}b (in Fig.~\ref{fig10}b3 we show only the attractor containing $O^+$).

The next important bifurcation occurs on the curve $PF_s$, where the pair of non-symmetric saddle fixed point merges with the symmetric stable one. After this, the symmetric fixed point becomes saddle, while the pair of two-winged Simo angels is still separated, see Fig.~\ref{fig10}c. They merge into one four-winged attractor after another heteroclinic bifurcation, when all four unstable separatrices $\Gamma^{\pm}_{1,2}(O^\pm)$ get to the stable manifold of the symmetric saddle periodic orbit $C_0$, see Fig.~\ref{fig10}d. As in the previous case, the attractor gets destroyed when a pair of stable and saddle fixed points emerges via a saddle-node bifurcation, see Fig.~\ref{fig10}f (blue and red periodic orbits correspond to these fixed points in Fig.~\ref{fig10}f3).

\begin{figure}[tb]
\center{\includegraphics[width=0.8\linewidth]{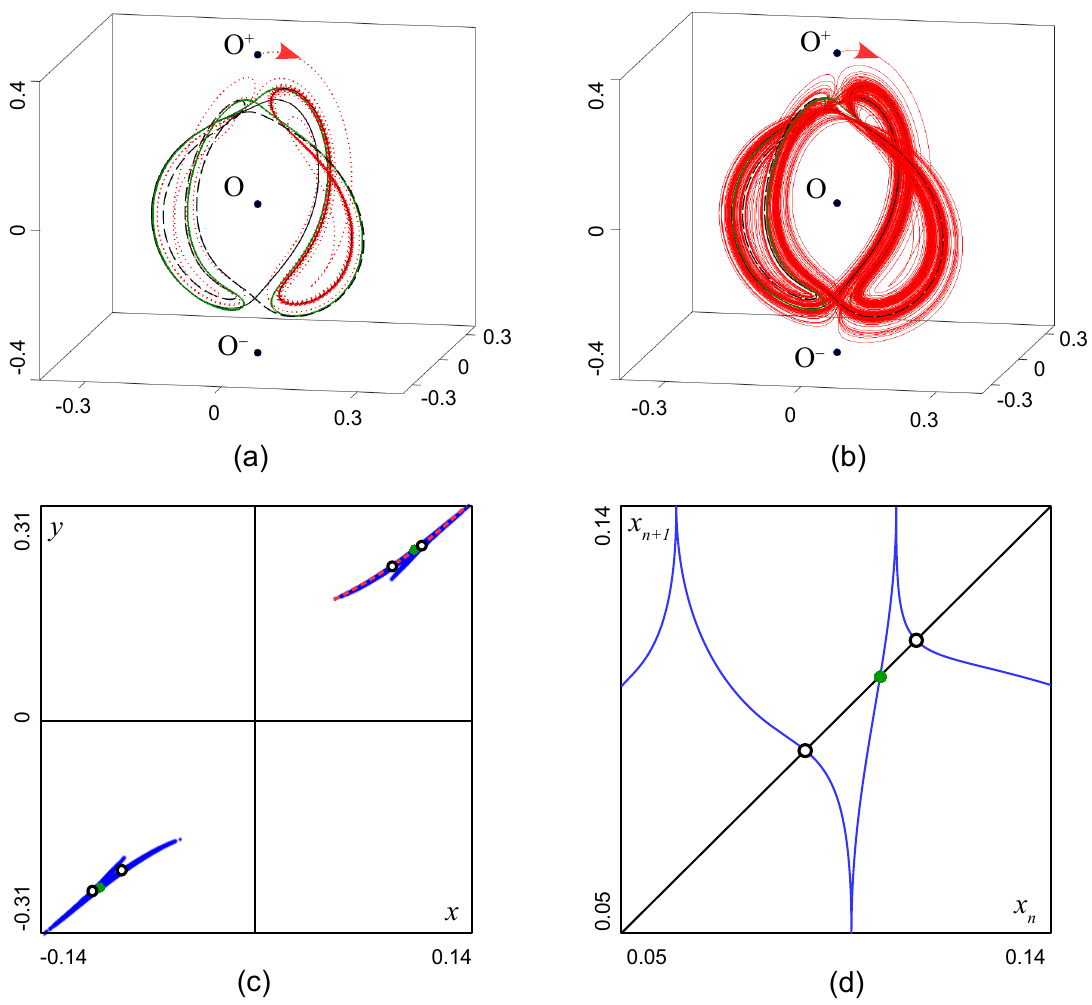} }
\caption{{\footnotesize Bifurcations along the pathway GH: $\beta=0.0075, \alpha \in [0.03, 0.05]$. (a) Phase portrait for $\alpha = 0.0039$: here we plot a pair of solid green periodic orbits and a pair of dashed black periodic orbits. (b) Phase portrait for $\alpha = 0.004$: Simo angel containing these four (saddle) periodic orbits; (c) and (d) corresponding 2D Poincar\'e and 1D first-return maps.}}
\label{fig11}
\end{figure}

\begin{remark}
The transition ``a symmetric fixed point $\to$ a pair of isolated chaotic attractors $\to$ one chaotic attractor'' is similar to the transition in the LA-region to the right of the curve $\nu = 1/2$ (when $1/2 < \nu < 1$) where the Lorenz attractor with a lacuna is born via a heteroclinic bifurcation (compare Figs.~\ref{fig5} and Figs.~\ref{fig10}).
\label{rem2}
\end{remark}

\begin{figure}[tb]
\center{\includegraphics[width=0.8\linewidth]{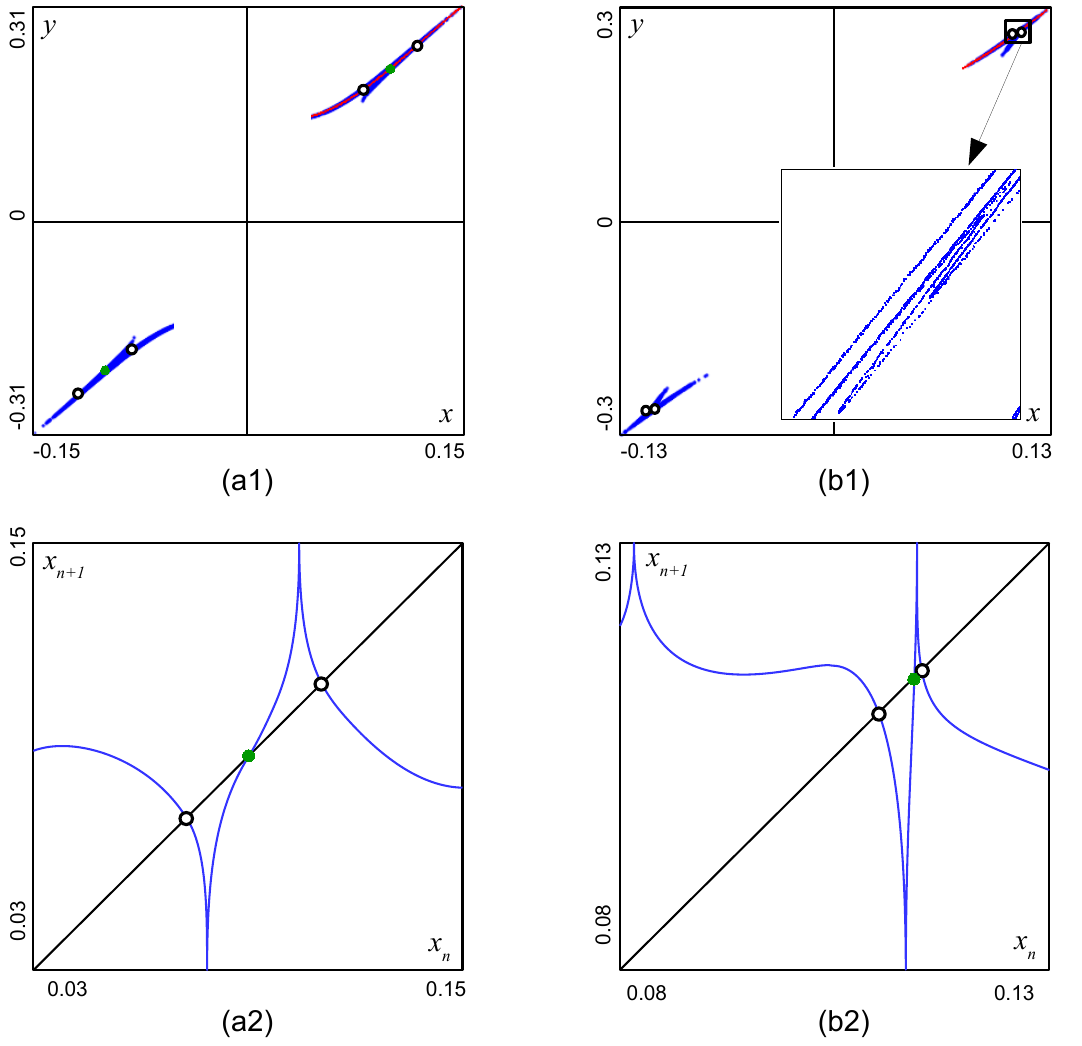} }
\caption{{\footnotesize Chaotic attractors at the points (a) I: $\alpha=0.04, \beta = 0.004$; (b) J: $\alpha=0.04, \beta = 0.011$. First row -- 2D first-return maps $T$ with the cross-section $z=0$, second row -- 1D first-return map constructed by the upper branch marked by red curve in the 2D maps.}}
\label{fig12}
\end{figure}

As we showed the both scenarios described above are similar to the scenarios leading to the birth of the Lorenz attractor inside the LA-region. However, crossing the curve $PD_s$, we observe a scenario which is different from those for the Lorenz attractor. Namely, consider the pathway GH: $\beta=0.0075, \alpha \in [0.03, 0.05]$. At the beginning, we observe four non-symmetric stable periodic orbits. The four separatrices $\Gamma^{\pm}_{1,2}(O^\pm)$ tend to these periodic orbits, each to its own, see Fig.~\ref{fig11}a. On the curve $PD_s$, these periodic orbits undergo the subcritical period-doubling bifurcation simultaneously and become a part of the four-winged Simo angel, see Fig.~\ref{fig11}b. The intersection of the attractor with the cross-section $z=0$ is shown in~Fig.~\ref{fig11}c. In addition to the four non-symmetric saddle periodic orbits (to pairs of white points), it also contains the symmetric periodic orbit $C_0$ (the green point). Note that the points of the attractor visibly do not lie on a smooth curve as it was in the previous cases. We consrtuct the 1D first-return map in this case only for the upper branch of the attractor marked by the red curve in Fig.~\ref{fig11}c. The resulting 1D map is shown in Fig.~\ref{fig11}d. One can see that this map is indeed close to the period-doubling bifurcation near the white points.

As we wrote above, not all attractors in the orange-colored domains of the SA-region (Fig.~\ref{fig7}) are pseudohyperbolic. In Figure~\ref{fig12} we show two examples of chaotic attractor found at points I: $\alpha=0.04, \beta = 0.004$ and J: $\alpha=0.04, \beta = 0.011$. In the first row we show the intersection of attractors with the cross-section $z=0$. In the second row we show the approximate 1D first-return maps. The critical points corresponding to the zero derivative in these 1D maps confirm that the attractors at the points I and J are not pseudohyperbolic.

\subsection{Numerical verification of pseudohyperbolicity} \label{sec3_3}

In this section we study pseudohyperbolic properties of the found Lorenz and Simo angel attractors. The pseudohyperbolicity conditions are much weaker than the hyperbolicity conditions. Nevertheless, they ensure the existence of a positive maximal Lyapunov exponent for all orbits in the attractor and the robustness of this property with respect to small perturbations (changes in the parameter values). Therefore, the verification of the pseudohyperbolicity of attractors takes the first priority in the general problem of studying their dynamics.

System \eref{eq_mainEq} is three-dimensional; for all its chaotic attractors we have one positive, one zero, and one strongly negative numerically evaluated Lyapunov exponents. Therefore, in the search of pseudohyperbolicity, we must assume that the invariant subspaces $E_1$ and $E_2$ in Def.~\ref{df_ph} are two- and one-dimensional, respectively. We call these subspaces central-unstable and strong-stable, and denote them $E^{cu}$ and $E^{ss}$. To verify the pseudohyperbolicity, we need to check that \cite{TS98, TS08}:
\begin{itemize}
    \item[{\rm (A)}] such splitting exists at every point of the attractor and depends continuously on the point;
    \item[{\rm (B)}] $E^{ss}$ corresponds to strong contraction (stronger than any possible contraction in $E^{cu}$);
    \item[{\rm (C)}] the differential of the system expands two-dimensional areas in $E^{cu}$.
\end{itemize}

Numerical methods for the verification of these conditions were developed in Refs.~\cite{Ginelli2007, Wolfe2007, Kup12, KupKuz2018, GKT21, GKKT21}. To check conditions (B) and (C), one just should calculate the spectrum of Lyapunov exponents and check that:
\begin{itemize}
    \item $\Lambda_2 > \Lambda_3$, which implies the fulfilment of condition (B);
    \item $\Lambda_1 + \Lambda_2 > 0$, which implies the fulfilment of condition (C).
\end{itemize}
Both these conditions are fulfilled for all chaotic attractors observed in system~\eref{eq_mainEq}.

Checking condition (A) is a much more difficult problem. It is based on the calculation of Lyapunov co-variant vectors~\cite{Kup12, KupKuz2018} and either constructing the so-called $E^{ss}$- and $E^{cu}$-continuity diagrams, as was proposed in Ref.~\cite{GKT21}, or computing the angles between $E^{ss}$ and $E^{cu}$, as was done in Ref.~\cite{Kup12, KupKuz2018}. Since in our case $dim E^{cu} = 2$, the continuity of $E^{cu}$ is equivalent to the continuity of the field of normals $N^{cu}$ to $E^{cu}$.

\begin{figure}[tb]
\center{\includegraphics[width=1.0\linewidth]{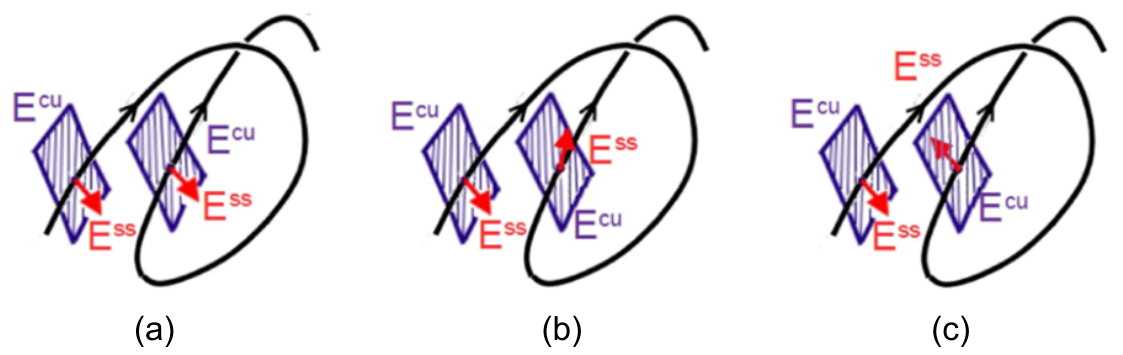} }
\caption{{\footnotesize Dependence of the subspaces $E^{ss}$ and $E^{cu}$ on the point for: (a) orientable pseudohyperbolic attractor; (b) not pseudohyperbolic attractor; (c) non-orientable pseudohyperbolic attractor.}}
\label{fig13}
\end{figure}

The $E^{ss}$-continuity diagram is a graph showing how angles $\varphi(v_i,v_j)$ between Lyapunov co-variant vectors $v_i \in E^{ss}(x_i)$ and $v_j \in E^{ss}(x_j)$ depend on the distance $\rho(x_i,x_j)$ between $x_i$ and $x_j$ for all pairs of points $x_i$ and $x_j$ in the attractor, see Ref.~\cite{GKT21}. Similarly, the $N^{cu}$-continuity diagram is built for vectors $w_j \in N^{cu}(x_j)$. One of the following three situations are possible for chaotic attractors (see Fig.~\ref{fig13}):
\begin{enumerate}
\item If $\varphi(v_i,v_j) \to 0$ and $\varphi(w_i,w_j) \to 0$ when $\rho(x_i,x_j) \to 0$, one can conclude that the subspaces $E^{ss}$ and $E^{cu}$ depend continuously on the point of the attractor (Fig.~\ref{fig13}a).
\item If, at $\rho(x_i,x_j) \to 0$, the angles $\varphi(v_i,v_j)$ or $\varphi(w_i,w_j)$ take arbitrary values, then the continuity is broken and the attractor is definitely non-pseudohyperbolic (Fig.~\ref{fig13}b).
\item If the values of $\varphi(v_i,v_j)$ or $\varphi(w_i,w_j)$ jump between $0$ and $\pi$  when $\rho(x_i,x_j) \to 0$ then the attractor is either pseudohyperbolic non-orientable (i.e., with a non-orientable field $E^{ss}$, like in the case of Simo angels, see Sec.~\ref{sec2}) or it is non-pseudohyperbolic (Fig.~\ref{fig13}c). To distinguish between these two possibilities one can compute the angles between the subspaces $E^{ss}$ and $E^{cu}$ at each point of the attractor: if the angles are bounded away from zero, this confirms the continuity of $E^{ss}$ and $E^{cu}$.
\end{enumerate}

Another method of checking the condition (A) is based on computing the angles between $E^{ss}$ and $E^{cu}$ at each numerically generated point $x_i$ in the attractor. If angles $\pi/2 - |\alpha(v_i,w_i)|$ are separated from zero at each point, then this is equivalent to the fulfilment of the $E^{ss}$- and $E^{cu}$-continuity conditions.

\begin{figure}[tb]
\center{\includegraphics[width=1.0\linewidth]{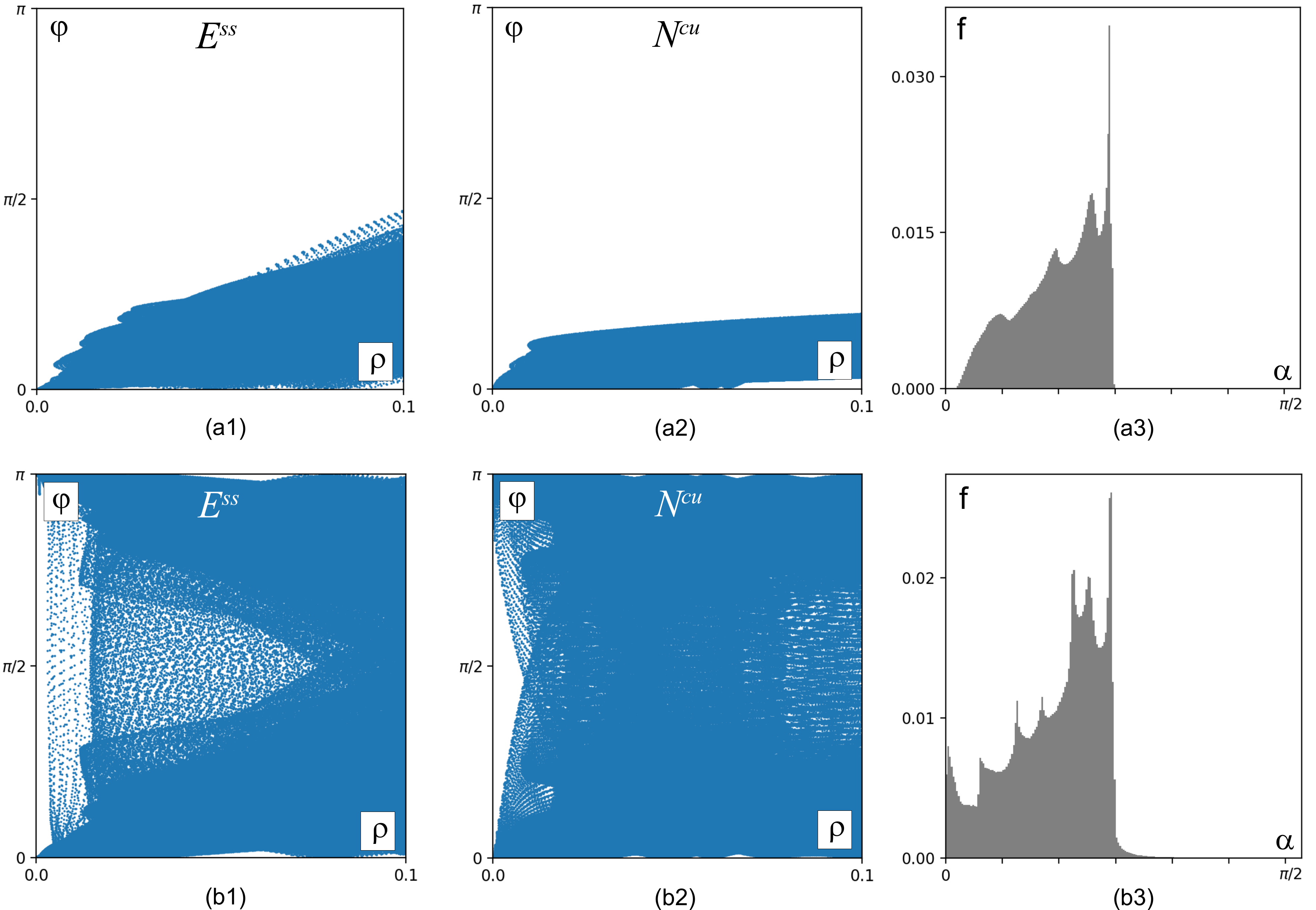} }
\caption{{\footnotesize Numerical verification of pseudohyperbolicity for the Lorenz-like attractors whose quotient maps are shown in Fig.~\ref{fig1}a ($\alpha = 0.1, \beta = 0.2$) -- first row and in Fig.~\ref{fig1}b ($\alpha = 0.092, \beta = 0.2$) -- second row. First and second columns -- the $E^{ss}$- and $N^{cu}$- continuity diagrams; third column -- histograms of angles between $E^{ss}$ and $E^{cu}$ for these attractors. The results confirm pseudohyperbolicity of the first attractor and non-pseudohyperbolicity for the second one.}}
\label{fig14}
\end{figure}

We apply both continuity and angle methods for the pair of Lorenz-like attractors whose quotient 1D maps are shown in Fig.~\ref{fig6}. The results are presented in Figure~\ref{fig14}: in the first row we check condition (A) for the attractor at $\alpha = 0.1, \beta = 0.2$, in the second row -- for the attractor at $\alpha = 0.092, \beta = 0.2$.

The $E^{ss}$- and $N^{cu}$-continuity diagrams shown in Figs.~\ref{fig14}a1, a2 confirm that the subspaces $E^{ss}$ and $E^{cu}$ depend continuously on the point of the first Lorenz-like attractor. The distribution of angles between $E^{ss}$ and $E^{cu}$ for this attractor is shown in Fig.~\ref{fig14}a3. The envelope curve does not touch the line $\alpha = 0$ of zero angles which confirms the absence of tangencies between the subspaces $E^{ss}$ and $E^{cu}$. Together these results allow us to conclude that the first attractor is pseudohyperbolic. The $E^{ss}$- and $N^{cu}$-continuity diagrams for the second attractor (Fig.~\ref{fig14}b1, b2) show that $\varphi(v_i,v_j)$ and $\varphi(w_i,w_j)$ can take arbitrary angles when $\rho(x_i,x_j) \to 0$. Moreover, the angles between these subspaces are not separated from zero. Therefore, we conclude that the second attractor is not pseudohyperbolic. This fact is in a good agreement with the graph of its 1D first return map (Fig.~\ref{fig6}b). The presence of a critical point with the zero derivative prevents the attractor to be pseudohyperbolic.

\begin{figure}[tb]
\center{\includegraphics[width=1.0\linewidth]{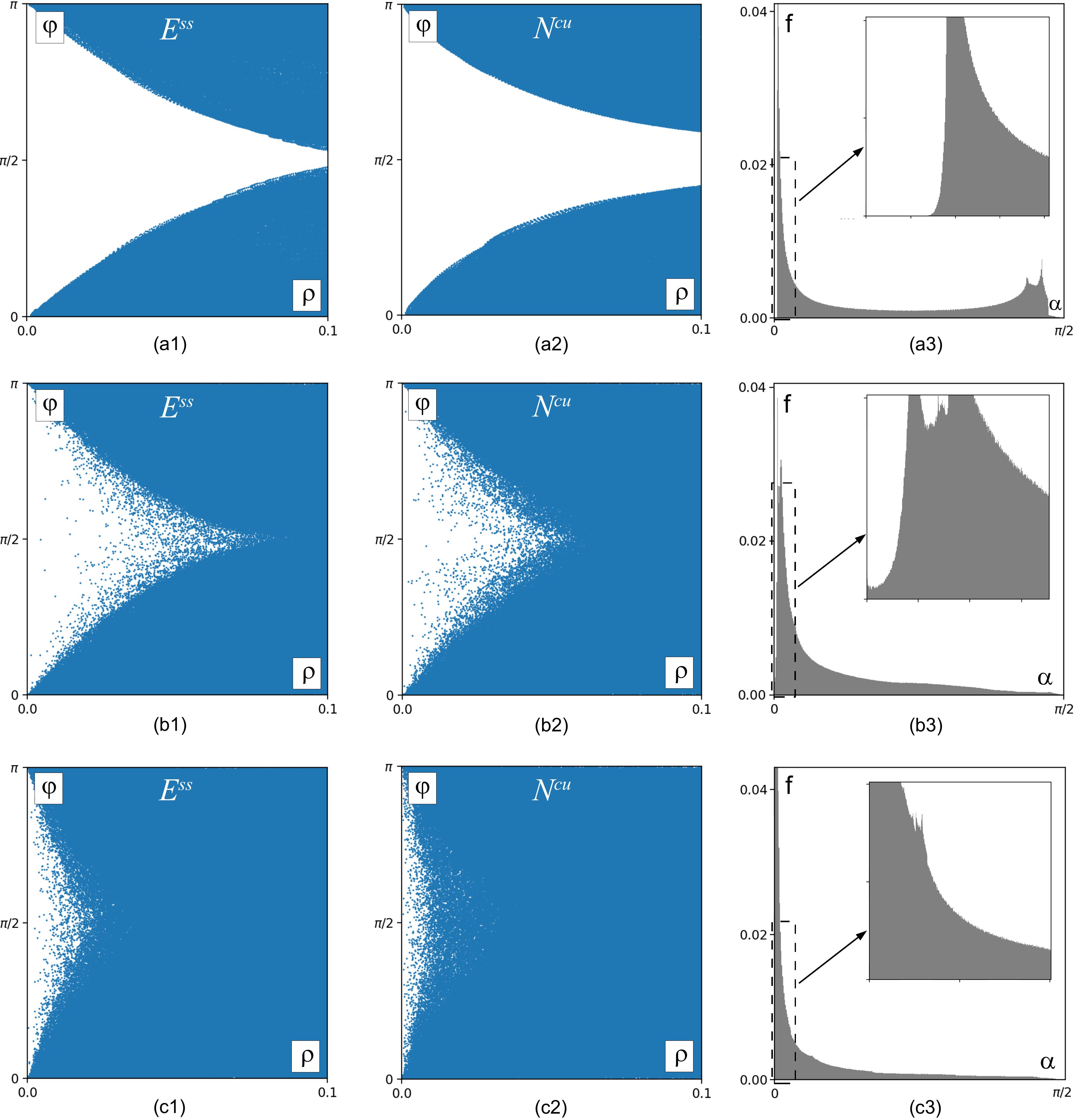} }
\caption{{\footnotesize Numerical verification of pseudohyperbolicity for the the Simo-like attractors shown in Fig.~\ref{fig8}a -- first row, and in Figs.~\ref{fig12}a and b -- second and third rows. First and second columns -- $E^{ss}$- and $N^{cu}$- continuity diagrams; third column -- histograms of angles between $E^{ss}$ and $E^{cu}$ for these attractors. The results confirm pseudohyperbolicity of the first attractor and non-pseudohyperbolicity for the second and third ones.}}
\label{fig15}
\end{figure}

Further, we check pseudohyperbolicity condition (A) for different types of Simo angels. For the attractor shown in Fig.~\ref{fig8}a the corresponding graphs are presented in the first row of Fig.~\ref{fig15}. The $E^{ss}$- and $N^{cu}$-continuity diagrams (Figs.~\ref{fig15}a1, a2) show that here we have the third case described above, i.e., the attractor may be either non-orientable pseudohyperbolic or not pseudohyperbolic. In Sec.~\ref{sec2} we show that the first-return map $T$ for the Simo angels reverses orientation in $E^{ss}$. Therefore, we conclude that the results of $E^{ss}$- and $N^{cu}$-continuity verification (Figs.~\ref{fig15}a1, a2) support the pseudohyperbolicity of the observed attractor. Additionally we calculate the histogram of angles between $E^{ss}$ and $E^{cu}$. This graph is shown in Fig.~\ref{fig15}a3. The absence of zero angles also confirms the pseudohyperbolicity of the attractor shown in Fig.~\ref{fig8}a. In the second and third rows of Fig.~\ref{fig15}, we analyze the fulfilment of condition (A) for the attractors presented in Fig.~\ref{fig12}a and ~\ref{fig12}b, respectively. Here, the $E^{ss}$- and $N^{cu}$-continuity diagrams (first and second columns of Fig.~\ref{fig15}) as well as the histogram of angles (third column of Fig.~\ref{fig15}) show that both these attractors are not pseudohyperbolic.

\begin{figure}[tb]
\center{\includegraphics[width=1.0\linewidth]{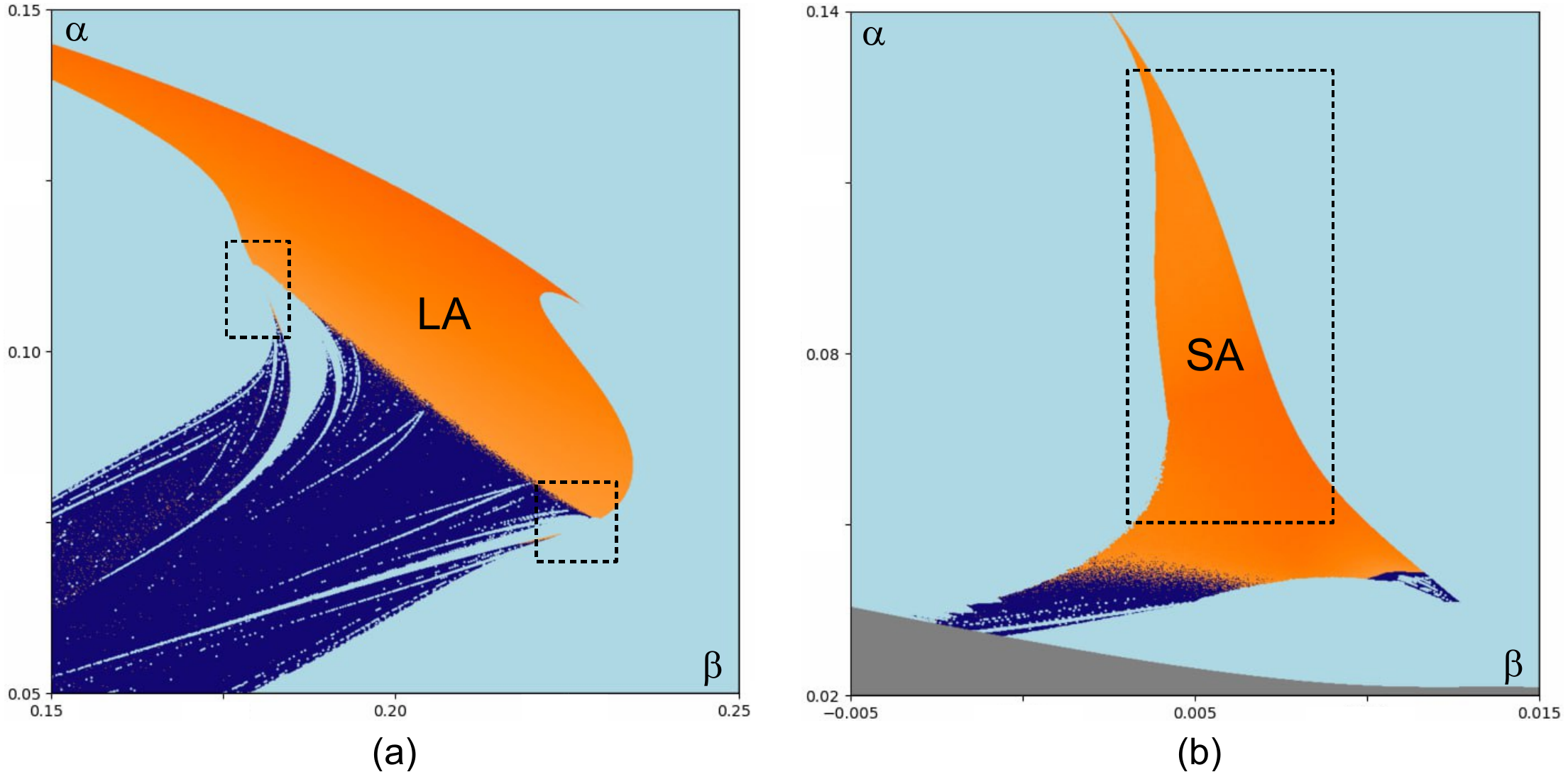} }
\caption{{\footnotesize Diagrams of minimal angles between $E^{ss}$ and $E^{cu}$ on the $(\beta, \alpha)$-parameter plane of the normal form \eref{eq_mainEq} for (a) region with Lorenz-like chaotic attractors; (b) region with Simo-like attractors. In the regions colored in dark blue the minimal angle estimated along an orbit of length $10^5$ is less than 0.01.}}
\label{fig16}
\end{figure}

Finally, for determining regions with pseudohyperbolic attractors in the $(\beta, \alpha)$-parameter plane we calculate minimal angle between $E^{ss}$ and $E^{cu}$ for chaotic attractors in the LA- and SA-regions shown in Fig.~\ref{fig4} and in Fig.~\ref{fig7}a, respectively. If the minimal angle estimated along an orbit of length 100000 is less than 0.01, we decide that the attractor is not pseudohyperbolic and color the corresponding pixel in dark blue, other regions with chaotic dynamics (a positive numerically evaluated Lyapunov exponent and the minimal angle between $E^{ss}$ and $E^{cu}$) are colored in orange, i.e., the orange regions correspond to pseudohyperbolic attractors. The corresponding diagrams are shown in Figure~\ref{fig16}. The results obtained for the region with the Lorenz-like chaotic attractors (Fig.~\ref{fig16}a) are in a good agreement with the results for the Shimizu-Morioka system~\cite{ASh93}. Here, pseudohyperbolic attractors exist mainly above the curve $l_{A=0}$. However thin regions with non-orientable pseudohyperbolic attractors -- the so-called Shilnikov flames \cite{AShil2014} -- also exist below this curve. For the region with Simo angels (Fig.~\ref{fig16}b) there are two large regions where the pseudohyperbolicity is violated. It looks like the left part of the boundary between pseudohyperbolic attractors and quasiattractors is similar to those in the Lorenz system where it is formed exactly by the curve $l_{A=0}$. The organization of the right part of this boundary is unclear for us. We will return to this question in future studies.

\subsection{Structure of homoclinic and heteroclinic bifurcations} \label{sec_3_4}

\begin{figure}[tb]
\center{\includegraphics[width=1.0\linewidth]{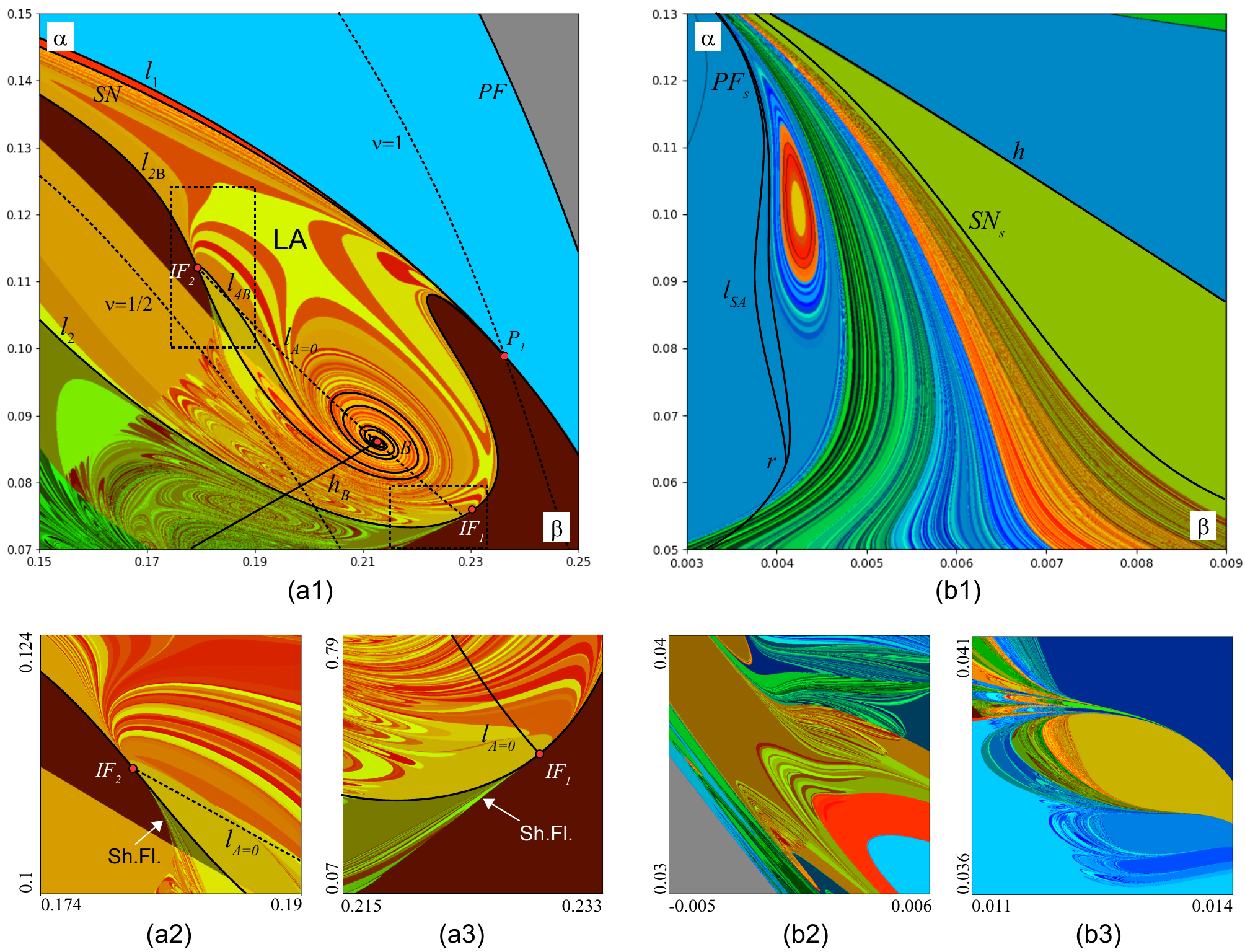} }
\caption{{\footnotesize Kneading diagrams for the normal form \eref{eq_mainEq} in the $(\beta, \alpha)$-parameter plane. Panel (a) shows the nice-foliated structure of the diagram inside the LA region. Panel (b) shows the nice-foliated structure of the diagram inside the SA region. Regular structure of kneading diagrams inside the regions LA and SA additionally confirms that the corresponding attractors satisfy (a) Afraimovich-Bykov-Shilnikov geometric model~\cite{ABS77, ABS82} and (b) geometric model described in Sec.~\ref{sec2}.}}
\label{fig17}
\end{figure}

It is known that the existence region of the Lorenz attractor in the parameter plane in the Shimizu-Morioka and Lorenz systems is foliated by curves corresponding to homoclinic butterflies \cite{ABS82, AShil2012, AShil2014}. In \cite{AShil2012, AShil2014, AShil18, AShil21}, a fast and effective method for computing such foliations -- the so-called \textit{kneading diagrams} -- was proposed.

In this section we employ this method for the study of homoclinic bifurcations inside and outside the region LA. We have also adapted this method for computing diagrams of heteroclinic connections between the pair of equilibria $O^+$ and $O^-$, and employ it to find heteroclinic connections inside the region SA. First, let us explain the essence of this method and, then, demonstrate and discuss the resulting kneading diagrams.

As in \cite{AShil2012, AShil2014}, one computes the kneading diagram in the following way. For any given parameter value, take the unstable separatrix $\Gamma^+_1(O^+)$ and use it to build the kneading sequence $s_0,s_1,s_2, \dots$ (by the symmetry $\mathcal{S}$, computations starting with any other of three separatrices will give symmetrical results). If, on this separatrix, the first point corresponding to the maximum of $|x|$ gives $x>0$, then put $s_0=1$, and if the first maximum of $|x|$ corresponds to $x<0$, then put $s_0=0$. Repeat this procedure and compute the numbers $s_j$ equal to $0$ or $1$ for $j=1,\dots, q$, where $q$ is any aforehand given integer. In fact, since we always take the same separatrix $\Gamma^+_1$, one always has $s_0=1$, so $s_0$ can be removed from the sequence.

Then, for each kneading segment $(s_1, s_2, \dots, s_q)$ one defines the value $D = \sum_{i=1}^q s_i 2^{q-i}$. Note that $D$ can take any integer values in the range $[0, 2^{q} - 1]$, and two kneading segments of the length $q$ are equal if and only if the corresponding values $D$ are equal. This means that the boundaries in the parameter plane between regions with different values of $D$ correspond to the appearance of a homoclinic loop. For the visualization of these boundaries, one assigns a different color to different values of $D$. The resulting picture is the kneading diagram. By construction, the change in color on the parameter plane indicates the change in the kneading, hence -- a homoclinic butterfly bifurcation curve. In order to obtain contrasting color pictures, each integer value from $[0, 2^q - 1]$ is converted to the RGB colors using the scheme proposed in \cite{AShil2014}. The values of $D$ from the segment $[0, (2^q - 1)/2)$ are converted to the volumes of the red channel, while the blue channel has intensity $0$. The values of $D \in [(2^q - 1)/2, 2^q - 1)]$ are converted to the volumes of the blue channel, while the red channel has intensity $0$. In both cases the volume of the green channel takes a random value. We are grateful to Andrey Shilnikov who explained this important know-how to us. We also grateful to Vladislav Koryakin who computed the kneading diagrams described in this section.

In Figure~\ref{fig17}a we show a kneading diagram with $q=16$ for the region with Lorenz-like attractors shown in Fig.~\ref{fig16}a. For the convenience we superimposed it with several obtained by MatCont~\cite{dhooge2008new, de2012interactive} bifurcation curves described in Sec.~\ref{sec3_1}. From this figure one can see that:
\begin{enumerate}
\item All homoclinic bifurcation curves found by the MatCont package are well fitted to the colored kneading diagram, which confirm the applicability of the method of kneading diagrams;
\item The kneading diagram is very similar to the kneading diagram for the Shimizu-Morioka system \cite{AShil2014}, which is consistent with the fact that system \eref{eq_mainEq} unfolds the codimension three bifurcation when the equilibria $O^+$ and $O^-$ have three zero eignevalues \cite{KKST24};
\item Like in the Shimizu-Morioka system, inside the region LA homoclinic bifurcation curves form a regular foliation with a same set of saddle singularities;
\item Also like in the Shimizu-Morioka system, below the curve $l_{A=0}$, the kneading diagram loses regularity (becomes blurred) which indicates that the chaotic attractor stops to be pseudohyperbolic due to the appearance of homoclinic tangencies \cite{GKT21, AShil2014}. The regularity of the foliation is restored inside thin regions, the so-called Shilnikov flames, which correspond to non-orientable (pseudohyperbolic) Lorenz attractors, see Fig.~\ref{fig17}a2, a3. These regions go out from the inclination-flip points \cite{GolHom11, ASh93, SST93};
\item We have three visible codimension-2 points from which the boundaries of the LA region start: two inclination-flip bifurcations $IF_{1,2}$ and the homoclinic butterfly with a neutral saddle $P_1$ (the fourth point $P_2$ with the four-winged heteroclinic connection is not shown in the figure).
\end{enumerate}

The kneading diagram for the region of the existence of the Simo angels is shown in Figure~\ref{fig17}b, it corresponds to the rectangular region of Fig.~\ref{fig16}b. As one can see, inside this region the heteroclinic bifurcation curves form a regular foliation which has the same structure as the kneading diagram for the 1D Lorenz map $x \mapsto |1-c x^\nu|$ on the plane of the parameters $(c, \nu)$ computed in~\cite{MS21}. This additionally confirms the pseudohyperbolicity of the corresponding attractors.

In Figs.~\ref{fig17}b2, c2 we present kneading diagrams in regions shown in Figs.~\ref{fig7}b and c, respectively. The blurred structure of these diagrams suggests the appearance of homoclinic tangencies and the destruction of the pseudohyperbolicity of the corresponding attractors.

\section{Pseudohyperbolic attractors in the three-dimensional H\'enon map} \label{sec4}

In this section we study chaotic attractors in the three-dimensional H\'enon map \eref{eq_HenonMap} near the codimension-3 point $(M_1, M_2, B) = (7/4,-1,-1)$ where this map has multipliers $(-1, i, -i)$. Let us shift the fixed point $P$ (the one which gives rise to chaotic attractors) to the origin. The resulting map has the following form:
\begin{equation}
\left\{
\begin{array}{l}
\bar x = y, \\
\bar y = z, \\
\bar z = Bx + Ay - Cz - z^2.
\end{array}
\right.
\label{eq_HenonMap0}
\end{equation}
Its zero fixed point $O_0$ has multipliers $(-1, i, -i)$ when $(C,A,B) = (1,-1,-1)$.

\begin{figure}[tb]
\center{\includegraphics[width=1.0\linewidth]{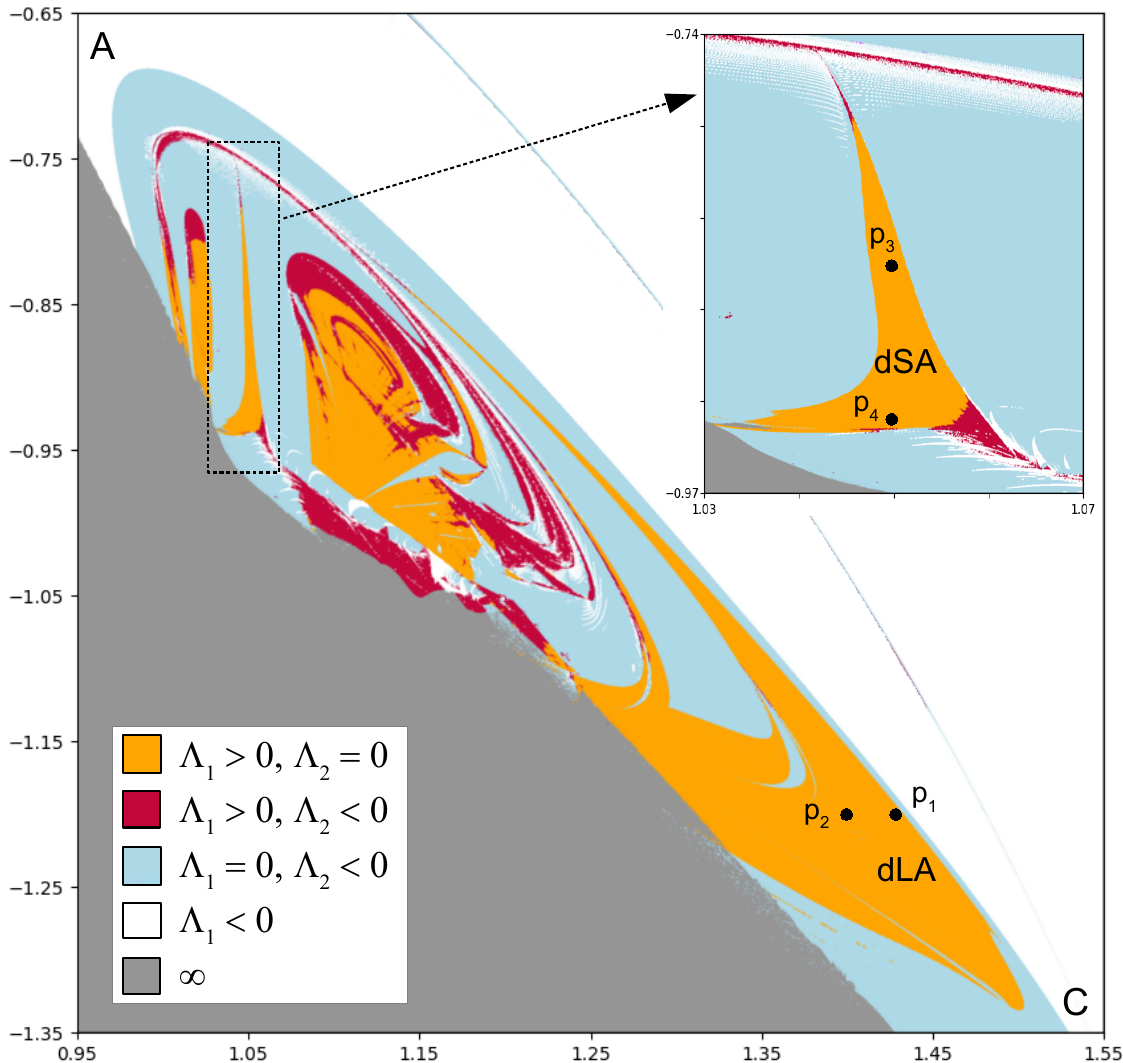} }
\caption{{\footnotesize Lyapunov diagram for map \eref{eq_HenonMap0} on the $(C,A)$-parameter plane. The parameter $B$ is determined by the parameters $A$ and $C$ according to the relation \eref{eq_B} with $\mu = 0.04$ (as in the normal form \eref{eq_mainEq}). We use here the same colors as for system \eref{eq_mainEq} complementing them by the additional crimson color for ``strongly dissipative'' chaotic attractors with $\Lambda_1 > 0, \Lambda_2 < 0$, see the palette in the left-bottom corner. The threshold of a zero Lyapunov exponent is 0.002. dLA and dSA regions with discrete Lorenz and Simo angel attractors. Pseudohyperbolicity of attractors at the points $p_1, p_2, p_3$, and $p_4$ are studied in detail (see Figs.~\ref{fig19} and \ref{fig20}).}}
\label{fig18}
\end{figure}

Recall that when studying the normal form \eref{eq_mainEq} we fixed parameter $\mu$ and performed two-parameter analysis in the $(\beta, \alpha)$ plane. It is shown in \cite{KKST24} that a constant value of $\mu$ in the normal form corresponds to the following relation between the coefficients of map \eref{eq_HenonMap0}:
\begin{equation}
B = e^\mu A + e^{2 \mu} C - e^{3 \mu}.
\label{eq_B}
\end{equation}
Therefore, in order to compare the results with the analysis of system \eref{eq_mainEq}, we will perform the numerical study of map \eref{eq_HenonMap0} when the Jacobian $B$ satisfies the relation \eref{eq_B} with $\mu=0.04$ and $(C, A)$ running around the point $(1, -1)$.

First, we compute the Lyapunov diagram on the parameter plane $(C,A)$, see Figure~\ref{fig18}. Here we use the same color scheme as for system \eref{eq_mainEq} supplementing it with the additional crimson color for ``strongly dissipative'' chaotic attractors with $\Lambda_1 > 0, \Lambda_2 < 0$, see the palette in the left-bottom corner.

One can see that this diagram is very similar to the Lyapunov diagram for the normal form \eref{eq_mainEq}, cf. Fig.~\ref{fig18} and Fig.~\ref{fig1}. Here one can see regions dLA and dSA similar to the regions LA and SA of Fig.~\ref{fig1} which correspond to discrete Lorenz-like and discrete Simo angel chaotic attractors. It is interesting to note that for the most part of these regions the chaotic attractors are ``flow-like'': the middle Lyapunov exponent is indistinguishable from zero in the numerical experiments. This confirms the closeness of a certain iteration of our map (the fourth iteration in this case) to the time-1 map of the flow of a system of differential equations (the normal form \eref{eq_mainEq}).

As one can see, large parts of the regions dLA and dSA are free from stability windows which allows us to assume that the corresponding chaotic attractors are robust, hence pseudohyperbolic. The conditions (B) and (C) of the pseudohyperbolicity definition~\ref{df_ph} are fulfilled here, since $\Lambda_1 > 0, \Lambda_2 \approx 0$ and $\Lambda_3 \ll 0$. It only remains to check the first condition (A) of the continuity of the subspaces $E^{ss}$ and $E^{cu}$.

\begin{figure}[tbh]
\center{\includegraphics[width=1.0\linewidth]{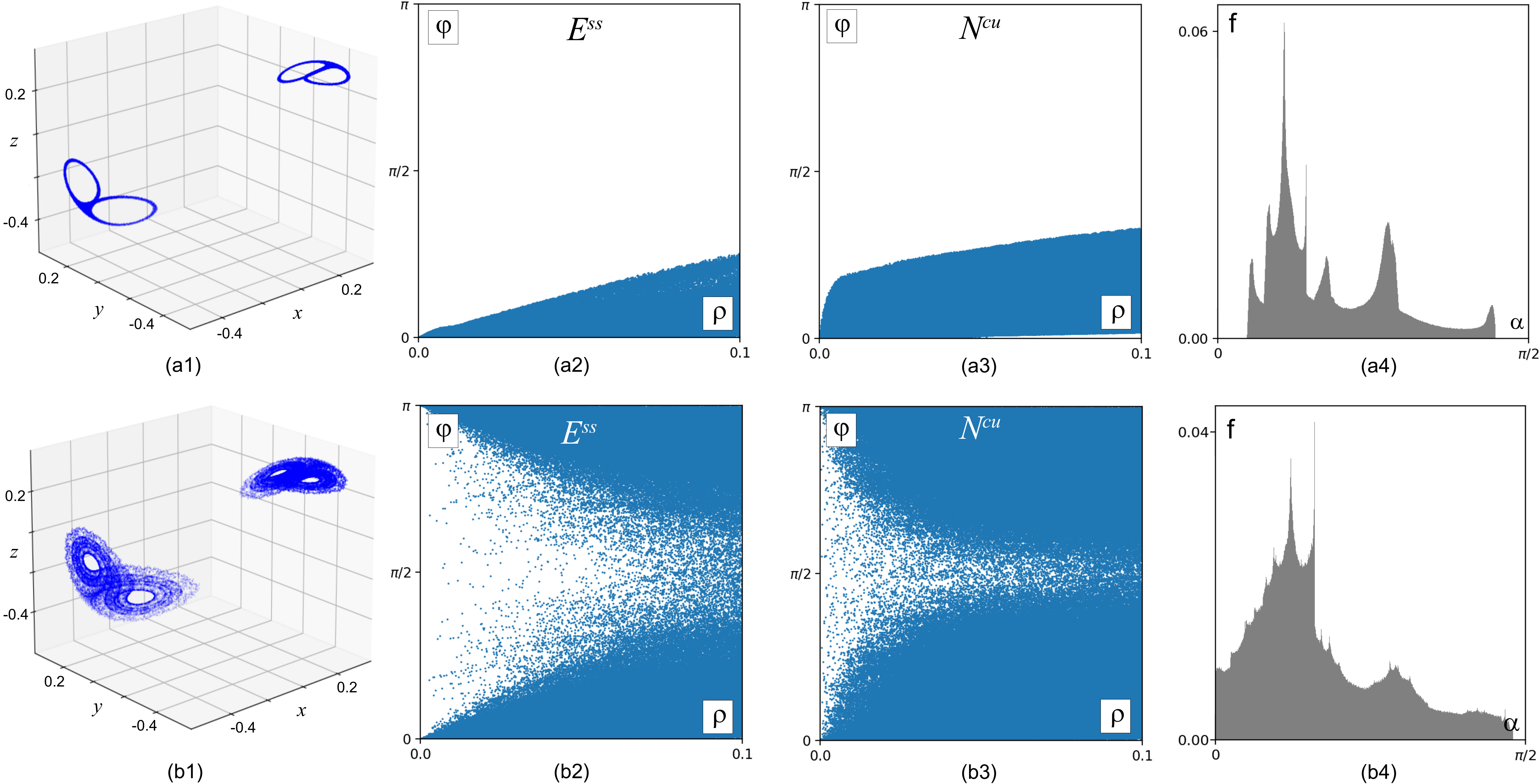} }
\caption{{\footnotesize Numerical verification of pseudohyperbolicity for the discrete Lorenz-like attractors existing at: $(C, A) = (1.43, -1.2)$ (point $p_1$ in Fig.~\ref{fig18}) -- first row and $(C, A) = (1.40, -1.2)$ (point $p_2$) -- second row. First column -- the corresponding phase portrait; second and third columns -- $E^{ss}$- and $N^{cu}$- continuity diagrams; fourth column -- histograms of angles between $E^{ss}$ and $E^{cu}$ for these attractors. The results confirm pseudohyperbolicity of the first attractor and non-pseudohyperbolicity for the second one.}}
\label{fig19}
\end{figure}

We check these conditions for the discrete Lorenz-like attractors at the parameter values $p_1$ and $p_2$ on the Lyapunov diagram. The point $p_1$ corresponds to the first row of Fig.~\ref{fig19} and $p_2$ corresponds to the second row. The phase portraits of the attractors are given in the first column of Fig.~\ref{fig19}. They look similar to the attractor of system \eref{eq_mainEq} shown in Fig.~\ref{fig0b}a, with the difference that in system \eref{eq_mainEq} we have a pair of Lorenz attractors symmetric to each other, whereas for map~\eref{eq_HenonMap0} this is one period-2 attractors (the pair of saddle equilibria of the normal form \eref{eq_mainEq} becomes a saddle period-2 orbit for map \eref{eq_HenonMap0}).

The pseudohyperbolicity check is shown in Figs.~\ref{fig19}a2--a4 and b2--b4. The subspaces $E^{ss}$ and $N^{cu}$ depend continuously on a point of the first attractor (the parameter point $p_1$; Fig.~\ref{fig19}a2, a3). The minimal angle between $E^{ss}$ and $E^{cu}$ for this attractor is clearly separated from zero (Fig.~\ref{fig19}a4). This confirm pseudohyperbolicity of the first discrete Lorenz attractor. On the other hand, the $E^{ss}$- and $N^{cu}$- continuity conditions are violated for the second attractor (the parameter point $p_2$; Fig.~\ref{fig19}b2, b3), and the minimal angle between $E^{ss}$ and $E^{cu}$ vanishes (Fig.~\ref{fig19}b4). Therefore, the second attractor cannot be pseudohyperbolic, even though it looks quite ``Lorenz-like''. Indeed, the point $p_2$ lies outside the pseudohyperbolicity region of Fig.~\ref{fig21}.

\begin{figure}[tb]
\center{\includegraphics[width=1.0\linewidth]{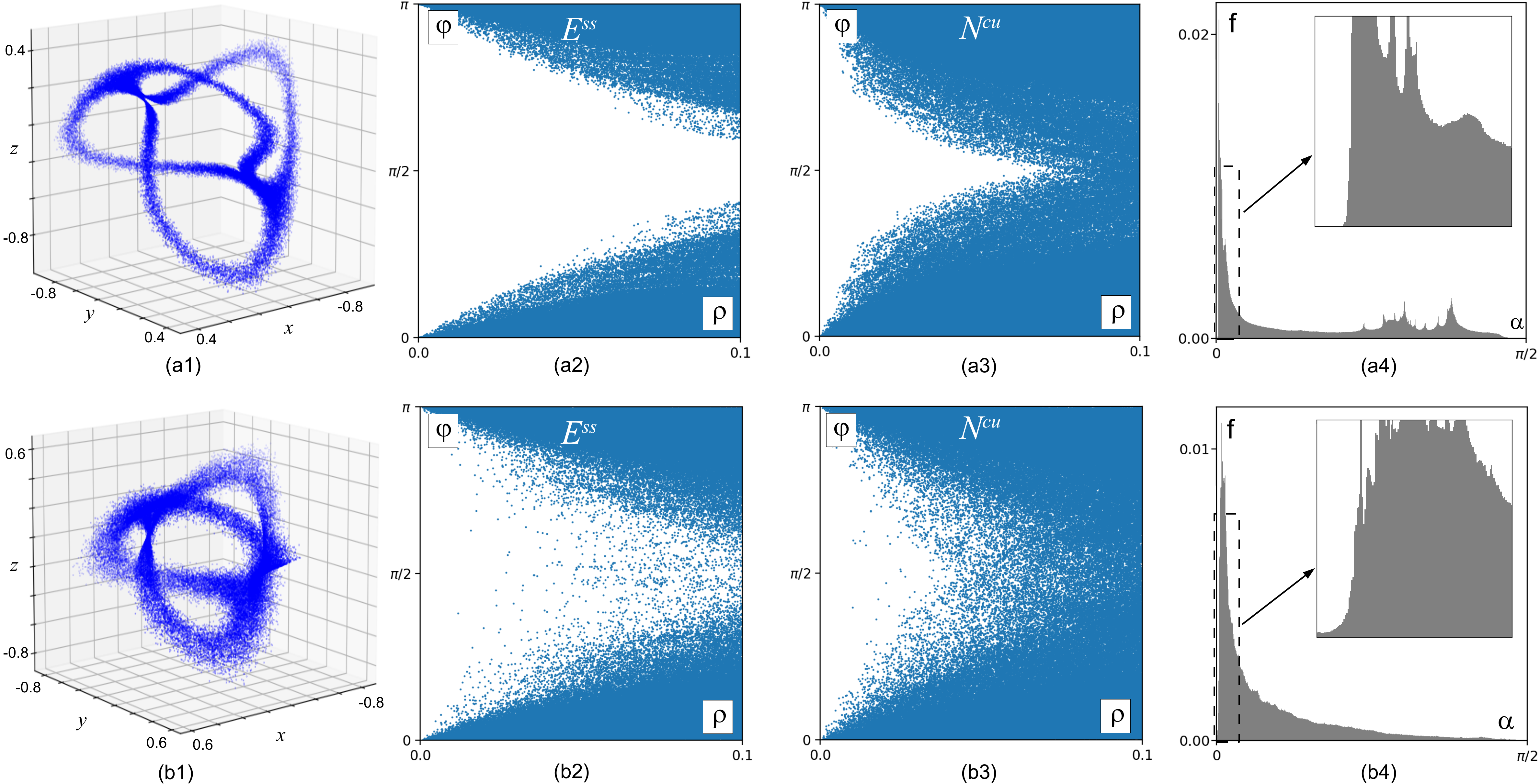} }
\caption{{\footnotesize Numerical verification of pseudohyperbolicity for the discrete Simo-like attractors existing at: $(C, A) = (1.05, -0.85)$ (point $p_3$ in Fig.~\ref{fig18}) -- first row and $(C, A) = (1.05, -0.93)$ (point $p_4$) -- second row. First column -- the corresponding phase portrait; second and third columns -- $E^{ss}$- and $N^{cu}$- continuity diagrams; fourth column -- histograms of angles between $E^{ss}$ and $E^{cu}$ for these attractors. The results confirm pseudohyperbolicity of the first attractor and non-pseudohyperbolicity for the second one.}}
\label{fig20}
\end{figure}

\begin{figure}[tb]
\center{\includegraphics[width=1.0\linewidth]{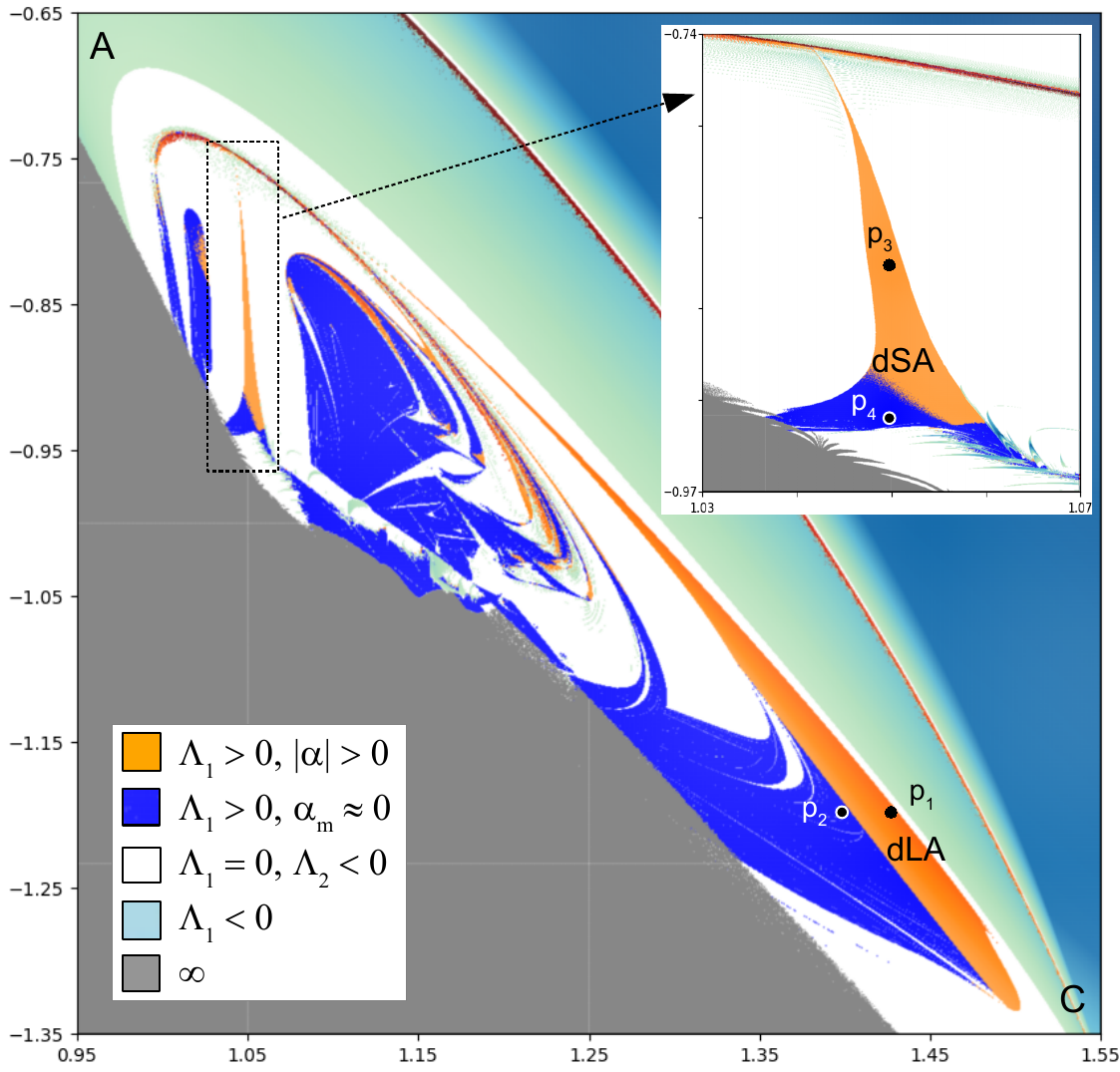} }
\caption{{\footnotesize Diagrams of minimal angles between $E^{ss}$ and $E^{cu}$ on the $(C, A)$-parameter plane of map \eref{eq_HenonMap0}. Inside regions colored in dark blue the minimal angle estimated along orbit of length 100000 is less than 0.001.}}
\label{fig21}
\end{figure}

The results of the $E^{ss}$- and $N^{cu}$-continuity verifications for the discrete Simo angels are shown in Figure~\ref{fig20}: in the first row we study the attractor at the parameter point $p_3$ (see its phase portrait in Fig.~\ref{fig20}a1), in the second row we verify the pseudohyperbolicity of the  attractor at the parameter point $p_4$ (its phase portrait is shown in Fig.~\ref{fig20}b1). The results confirm that the first attractor is non-orientable and pseudohyperbolic, whereas the second one cannot be pseudohyperbolic, since there are no continuity of the subspaces $E^{ss}$ and $E^{cu}$ and the tangencies between these subspaces are present.

In Figure~\ref{fig21}, we present results of the massive search for the pseudohyperbolic attractors in map \eref{eq_HenonMap0}. We augment the Lyapunov diagram by computations of the minimal angle between the subspaces $E^{ss}$ and $E^{cu}$ for chaotic attractors at each point of the parameter plane $(C,A)$. Regions where the absolute value of the minimal angle is less than a threshold value $0.001$ are colored in blue (the same as it was done for system \eref{eq_mainEq}). In these regions chaotic attractor are not expected to be pseudohyperbolic, in contrast to the orange colored regions where this angle is greater that $0.001$.

Here we observe two large pseudohyperbolicity regions dLA and dSA, practically the same as the regions LA and SA in the normal form \eref{eq_mainEq}. One can also see a sequence of pseudohyperbolcity regions supposedly adjoining the meeting point of the regions dLA and dSA. In the normal form \eref{eq_mainEq} this point corresponds to the four-winged heteroclinic connection. The theory of \cite{KKST24} indeed predicts that this bifurcation is an end point of a countable set of regions of existence of pseudohyperbolic Lorenz and Simo attractors. So, these regions confirm the theory. The origin of the other regions of pseudohyperbolicity in Fig.~\ref{fig21} is unknown to us.

\subsection*{Acknowledgment}

The authors are grateful to Sergey Gonchenko for useful discussions and Andrey Bobrovsky, Sergey Gonchenko, and Vladislav Koryakin, for the help with numerics. The work was supported by the Leverhulme Trust grant RPG-2021-072, the RSF grant No. 19-71-10048 (Sec.~\ref{sec2}), the RSF grant No. 23-71-30008 (Sec.~\ref{sec3}), and by the Basic Research Program at HSE (Sec.~\ref{sec4}).

\end{document}